\documentclass[aps,prl,preprint,tightenlines,superscriptaddress,showpacs,byrevtex]{revtex4}
\usepackage{graphicx} % Include figure files
\usepackage{dcolumn}  % Align table columns on decimal point

\graphicspath{{ps}}

\begin{document}

\preprint{\vbox{ \hbox{   }
                 \hbox{BELLE-CONF-0460}
                 \hbox{ICHEP04 11-0710} 
}}

\title{ \quad\\[0.5cm]  Study of $B^0\to \bar{D}^{(*)0} \pi^+ \pi^-$ decays}

\affiliation{Aomori University, Aomori}
\affiliation{Budker Institute of Nuclear Physics, Novosibirsk}
\affiliation{Chiba University, Chiba}
\affiliation{Chonnam National University, Kwangju}
\affiliation{Chuo University, Tokyo}
\affiliation{University of Cincinnati, Cincinnati, Ohio 45221}
\affiliation{University of Frankfurt, Frankfurt}
\affiliation{Gyeongsang National University, Chinju}
\affiliation{University of Hawaii, Honolulu, Hawaii 96822}
\affiliation{High Energy Accelerator Research Organization (KEK), Tsukuba}
\affiliation{Hiroshima Institute of Technology, Hiroshima}
\affiliation{Institute of High Energy Physics, Chinese Academy of Sciences, Beijing}
\affiliation{Institute of High Energy Physics, Vienna}
\affiliation{Institute for Theoretical and Experimental Physics, Moscow}
\affiliation{J. Stefan Institute, Ljubljana}
\affiliation{Kanagawa University, Yokohama}
\affiliation{Korea University, Seoul}
\affiliation{Kyoto University, Kyoto}
\affiliation{Kyungpook National University, Taegu}
\affiliation{Swiss Federal Institute of Technology of Lausanne, EPFL, Lausanne}
\affiliation{University of Ljubljana, Ljubljana}
\affiliation{University of Maribor, Maribor}
\affiliation{University of Melbourne, Victoria}
\affiliation{Nagoya University, Nagoya}
\affiliation{Nara Women's University, Nara}
\affiliation{National Central University, Chung-li}
\affiliation{National Kaohsiung Normal University, Kaohsiung}
\affiliation{National United University, Miao Li}
\affiliation{Department of Physics, National Taiwan University, Taipei}
\affiliation{H. Niewodniczanski Institute of Nuclear Physics, Krakow}
\affiliation{Nihon Dental College, Niigata}
\affiliation{Niigata University, Niigata}
\affiliation{Osaka City University, Osaka}
\affiliation{Osaka University, Osaka}
\affiliation{Panjab University, Chandigarh}
\affiliation{Peking University, Beijing}
\affiliation{Princeton University, Princeton, New Jersey 08545}
\affiliation{RIKEN BNL Research Center, Upton, New York 11973}
\affiliation{Saga University, Saga}
\affiliation{University of Science and Technology of China, Hefei}
\affiliation{Seoul National University, Seoul}
\affiliation{Sungkyunkwan University, Suwon}
\affiliation{University of Sydney, Sydney NSW}
\affiliation{Tata Institute of Fundamental Research, Bombay}
\affiliation{Toho University, Funabashi}
\affiliation{Tohoku Gakuin University, Tagajo}
\affiliation{Tohoku University, Sendai}
\affiliation{Department of Physics, University of Tokyo, Tokyo}
\affiliation{Tokyo Institute of Technology, Tokyo}
\affiliation{Tokyo Metropolitan University, Tokyo}
\affiliation{Tokyo University of Agriculture and Technology, Tokyo}
\affiliation{Toyama National College of Maritime Technology, Toyama}
\affiliation{University of Tsukuba, Tsukuba}
\affiliation{Utkal University, Bhubaneswer}
\affiliation{Virginia Polytechnic Institute and State University, Blacksburg, Virginia 24061}
\affiliation{Yonsei University, Seoul}
  \author{K.~Abe}\affiliation{High Energy Accelerator Research Organization (KEK), Tsukuba} % KEK
  \author{K.~Abe}\affiliation{Tohoku Gakuin University, Tagajo} % TohokuGakuin
  \author{N.~Abe}\affiliation{Tokyo Institute of Technology, Tokyo} % TIT
  \author{I.~Adachi}\affiliation{High Energy Accelerator Research Organization (KEK), Tsukuba} % KEK
  \author{H.~Aihara}\affiliation{Department of Physics, University of Tokyo, Tokyo} % Tokyo
  \author{M.~Akatsu}\affiliation{Nagoya University, Nagoya} % Nagoya
  \author{Y.~Asano}\affiliation{University of Tsukuba, Tsukuba} % Tsukuba
  \author{T.~Aso}\affiliation{Toyama National College of Maritime Technology, Toyama} % Toyama
  \author{V.~Aulchenko}\affiliation{Budker Institute of Nuclear Physics, Novosibirsk} % BINP
  \author{T.~Aushev}\affiliation{Institute for Theoretical and Experimental Physics, Moscow} % ITEP
% ...
% Lausanne

\collaboration{Belle Collaboration}

\noaffiliation

\begin{abstract}
 We report the results of a study of neutral $B$ decays to the $D^0 \pi^+ \pi^-$ and $D^{*0} \pi^+ \pi^-$ final states using complete $D^{(*)0}$ reconstruction. The contributions from two-body $B\to D^{**} \pi$ with narrow ($j_q=3/2$) $D^{**}$ states and $B\to D^{(*)}\rho, D^{(*)} f_2, D^{(*)}\sigma$ decays have been determined. All results are preliminary, and are based on a large data sample collected in the Belle experiment at the KEKB $e^+e^-$ collider.
\end{abstract}


\maketitle

\tighten

\begin{figure}[h]
\begin{center}
\begin{tabular}{c}
\includegraphics[height=9 cm, width=12 cm]{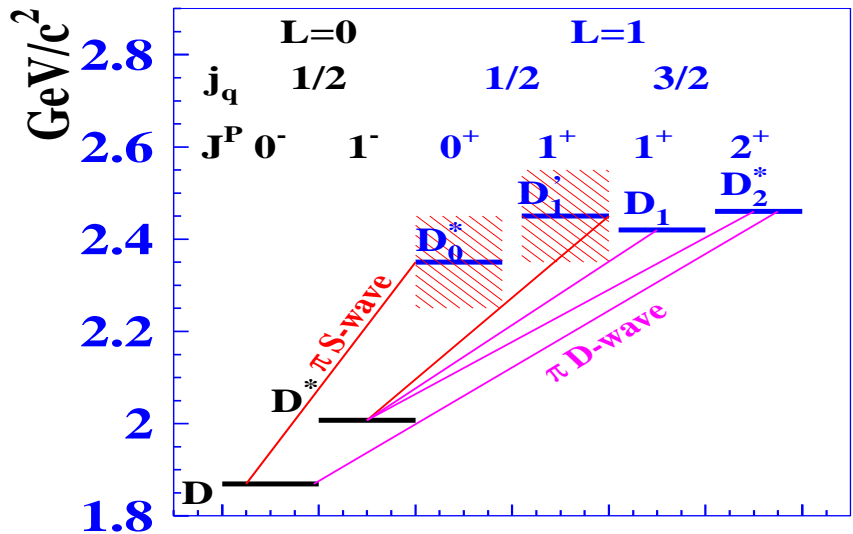}
\end{tabular}
\end{center}
\caption{Spectrum of D-meson excitations. Lines show possible one pion 
transitions.}
\label{fi:spec}
\end{figure}

{\renewcommand{\thefootnote}{\fnsymbol{footnote}}}
\setcounter{footnote}{0}
The decays of $B$ meson
to $D\pi$ and $D^*\pi$ final states are two of its 
dominant  hadronic decay modes and have been measured quite 
well~\cite{PDG}.
 In this paper we study the   production of excited states of $D$-mesons,
 collectively referred to as
$D^{**}$'s, that are P-wave excitations of quark-antiquark  systems 
containing one charmed and one light ($u,d$) quark. 
The results provide tests of Heavy Quark Effective Theory (HQET) and
QCD sum rules.
Figure~\ref{fi:spec} shows the spectrum of 
$D$-meson excitations. In the heavy quark limit, the heavy quark spin
${\vec s}_c$ decouples from the other degrees of freedom and the total
angular momentum of the light quark ${\vec j}_q=\vec{L}+{\vec s}_q$ is a good 
quantum number.
There are four P-wave states with the following spin-parity and light 
quark angular  momenta:
$0^+(j_q=1/2),~1^+(j_q=1/2),~1^+(j_q=3/2)$ and $2^+(j_q=3/2)$, which are 
usually labeled as $D^*_0,~D'_1,~D_1$ and $D^*_2$, respectively.   

The two $j_q=3/2$ states are narrow with widths of about 20-40~MeV
and were previously observed~\cite{AR1,AR2,AR3,e691,CL15,e687,CL2,dobs,DELPHI,DELPHI1,ALEPH}.
 The measured values of their masses 
agree with model predictions~\cite{isgur,rosner,godfrey,falk}.
The remaining $j_q=1/2$ states decay via S-waves and are expected to be 
quite broad.

The $B\to D^{(*)}\pi\pi$ decay provides a possibility to  study
$D^{**}$ production.  The fixed spin  of the  initial state makes
it possible to perform an angular analysis of the decay products 
and to separate final states with different quantum numbers.

We have acquired a large sample of $B$ decays using the Belle detector,
which has good resolution and particle ID.   These data can be 
used to contribute to an understanding of these problems.

Earlier neutral $D^{**}$ production in charged $B$-decays has been studied
at Belle~\cite{mybelle}. All four $D^{**}$ states have been
observed and the production rate of the broad ($j=1/2$)-states 
appears to be comparable to that of the narrow ($j=3/2$)-states.
This work describes a similar analysis of 
the decay $\bar{B}^0\to D^{**+}\pi^-$.

In the case of neutral $B$ decay the $D^{(*)}\pi\pi$ final state contains 
two pions of opposite sign  that can form several bound
states (such as $\rho,~f_0,~f_2$), which should also be taken into account.
The presence of $\pi\pi$ bound states complicates the analysis but 
can give valuable information about 
the constants and mechanism of these decays.

\section{The Belle detector}
  The Belle detector~\cite{Belle} is a large-solid-angle magnetic spectrometer
that consists
of a three-layer silicon vertex detector (SVD), 
a 50-layer central drift chamber
(CDC) for charged particle tracking and specific ionization measurement 
($dE/dx$), an array of aerogel threshold \v{C}erenkov counters (ACC), 
time-of-flight scintillation counters (TOF), and an array of 8736 CsI(Tl) 
crystals for electromagnetic calorimetry (ECL) located inside a superconducting
solenoid coil that provides a 1.5~T magnetic field. An iron flux return located
outside the coil is instrumented to detect $K_L$ mesons and identify muons
(KLM). 
We use a GEANT-based Monte Carlo (MC) simulation to
model the response of the detector and to determine its acceptance~\cite{sim}.

  Separation of kaons and pions is accomplished by combining the responses of 
the ACC and the TOF with $dE/dx$ measurements in the CDC
 to form a likelihood $\cal{L}$($h$) where $h=(\pi)$ or $(K)$. 
Charged particles 
are identified as pions or kaons using the likelihood ratio 
(${\mathcal R}$):
\[{\rm {\mathcal R}}(K)=\frac{{\cal{L}}(K)}{{\cal{L}}(K)+{\cal{L}}(\pi)};~~
{\rm {\mathcal R}}(\pi)=
\frac{{\cal{L}}(\pi)}{{\cal{L}}(K)+{\cal{L}}(\pi)}=1-{\rm {\mathcal R}}(K).\]
  At large momenta ($>$2.5 GeV/$c$) only the ACC and $dE/dx$ are used since the
TOF provides no significant separation of kaons and pions. 
Electron identification is based on a combination of $dE/dx$ measurements,
ACC photoelectron yields  and the position, shape and total energy deposition
($E/p$) of the shower detected in the ECL. A more detailed
description of the Belle particle identification can be found in 
Ref.~\cite{PID}.

\section{Event selection}

A data sample of 140~fb$^{-1}$ 
(152 million~$B\bar{B}$ events) collected at the 
$\Upsilon (4S)$ resonance with the Belle detector is used.
Candidate $\bar{B}^0\to D^0\pi^+\pi^-$ and $\bar{B}^0\to
D^{*0}\pi^+\pi^-$ 
 events
as well as charge conjugate combinations are selected.
The $D^0$ and $D^{*0}$ mesons are reconstructed in 
$D^0\to K^-\pi^+$ and $D^{*0}\to D^0\pi^0$ modes, respectively.
$D^0$ from $D^{*0}$ decay is detected in channels $D^0\to K^-\pi^+$ and
$D^0\to K^-\pi^+\pi^+\pi^-$.
The signal-to-background ratios for other $D$ decay modes are found to be much
lower and they are not used in this analysis.

Charged tracks are selected with requirements based on the  
average hit residuals and impact parameters relative to the interaction
point. We also require that the polar angle of each track be within
the angular  range of $17^{\circ}-150^{\circ}$ and that the transverse track 
momentum be 
greater than 50 MeV/$c$ for kaons and 25 MeV/$c$ for pions.

Charged kaon candidates are selected with the requirement ${\rm {\mathcal R}}(K)>0.6$. 
This  has an efficiency of
$90\%$ for kaons
and a pion misidentification probability of $10\%$.
For pions the requirement \mbox{${\rm {\mathcal R}}(\pi)>0.2$} is used.
All tracks that are positively identified as electrons are rejected. 

$D^{0}$ mesons are reconstructed from  
$K^{-}\pi^{+}$ combinations with
invariant mass within 12~MeV/c$^2$ of the nominal $D^0$ mass, which
corresponds to about 2.5\,$\sigma_{K\pi}$.
We reconstruct $D^{*0}$ mesons from the $D\pi^0$ combinations 
with a mass difference of $M_{D\pi^0}-M_{D^0}$ within $2.5~{\rm
  MeV}/c^2$ of its nominal value.

Candidate events are identified 
by their center of mass (c.m.)\ energy difference, 
$\Delta E=(\sum_iE_i)-E_{\rm b}$, and 
beam-constrained mass, 
$M_{\rm bc}=\sqrt{E^2_{\rm b}-(\sum_i\vec{p}_i)^2}$, where 
$E_{\rm b}=\sqrt{s}/2$ is the beam energy in the $\Upsilon(4S)$
c.m.\ frame, and $\vec{p}_i$ and $E_i$ are the c.m.\ three-momenta and 
energies of the $B$ meson candidate decay products. We select events with 
$M_{\rm bc}>5.25$~GeV/$c^2$ and $|\Delta E|<0.10$~GeV.

To suppress the large continuum background ($e^+e^-\to q\bar{q}$,
where $q=u,d,s,c$), topological variables are used. Since  
the produced $B$ mesons 
are almost at rest in the c.m. frame, the angles of the decay products
of the two $B$ mesons are uncorrelated and the tracks tend to be 
isotropic while  continuum $q\bar{q}$ events
tend to have  a two-jet structure. We use the angle between the thrust axis of 
the $B$ candidate and that of the rest of the event ($\Theta_{\rm thrust}$)
to discriminate between these two cases. The distribution of
$|\cos\Theta_{\rm thrust}|$ is strongly peaked near $|\cos\Theta_{\rm thrust}|=1$
for $q\bar{q}$ events and is nearly flat for  $\Upsilon(4S)\to
B\bar{B}$ events.
We require $|\cos\Theta_{\rm thrust}|<0.8$, which eliminates about 
83$\%$ of the continuum background while retaining about 80$\%$ of signal 
events.

There are events for which two or more combinations pass all
the selection criteria. According to a MC simulation,
this occurs  primarily
because of the misreconstruction of a low momentum pion from the 
$D^{**}\to D^{(*)}\pi$ decay. To avoid multiple entries, 
the combination that has the  minimum difference of 
$Z$ coordinates at the interaction point, $|Z_{\pi_1}-Z_{\pi_2}|$, 
of the tracks corresponding to the pions from 
$B\to D^{**}\pi_1$ and $D^{**}\to D^{(*)}\pi_2$ decays
is selected~\cite{foot1}.
 This selection 
suppresses the combinations that
include pions from $K_S$ decays. In the case of multiple $D$ 
combinations, the one with invariant
mass closest to the nominal value is selected.  
 
\section{$\bar{B}^{0}\to D^{0}\pi^{+}\pi^{-}$ analysis.}
The final state of the  $\bar{B}^{0}\to D^{0}\pi^{+}\pi^{-}$ decay 
together with three-body  and quasi-two-body events
includes  the two-body decay $\bar{B}^{0}\to D^{*+}\pi^{-}$ followed by the 
decay $D^{*+}\to D^{0}\pi^{+}$. 
Using the mass difference of $M_{D\pi}-M_{D}$ 
we subdivide the total sample in two: events with $|M_{D\pi}-M_{D}-0.1455|<0.03~\rm
(GeV/c^2)$($\sim6\sigma$) (denoted further
as sample (2)) correspond to $D^*\pi$ production and the rest of the events 
$D\pi\pi$ form sample (1).

\begin{figure}[h]
\begin{tabular}{cc}
 \includegraphics[height=8 cm]{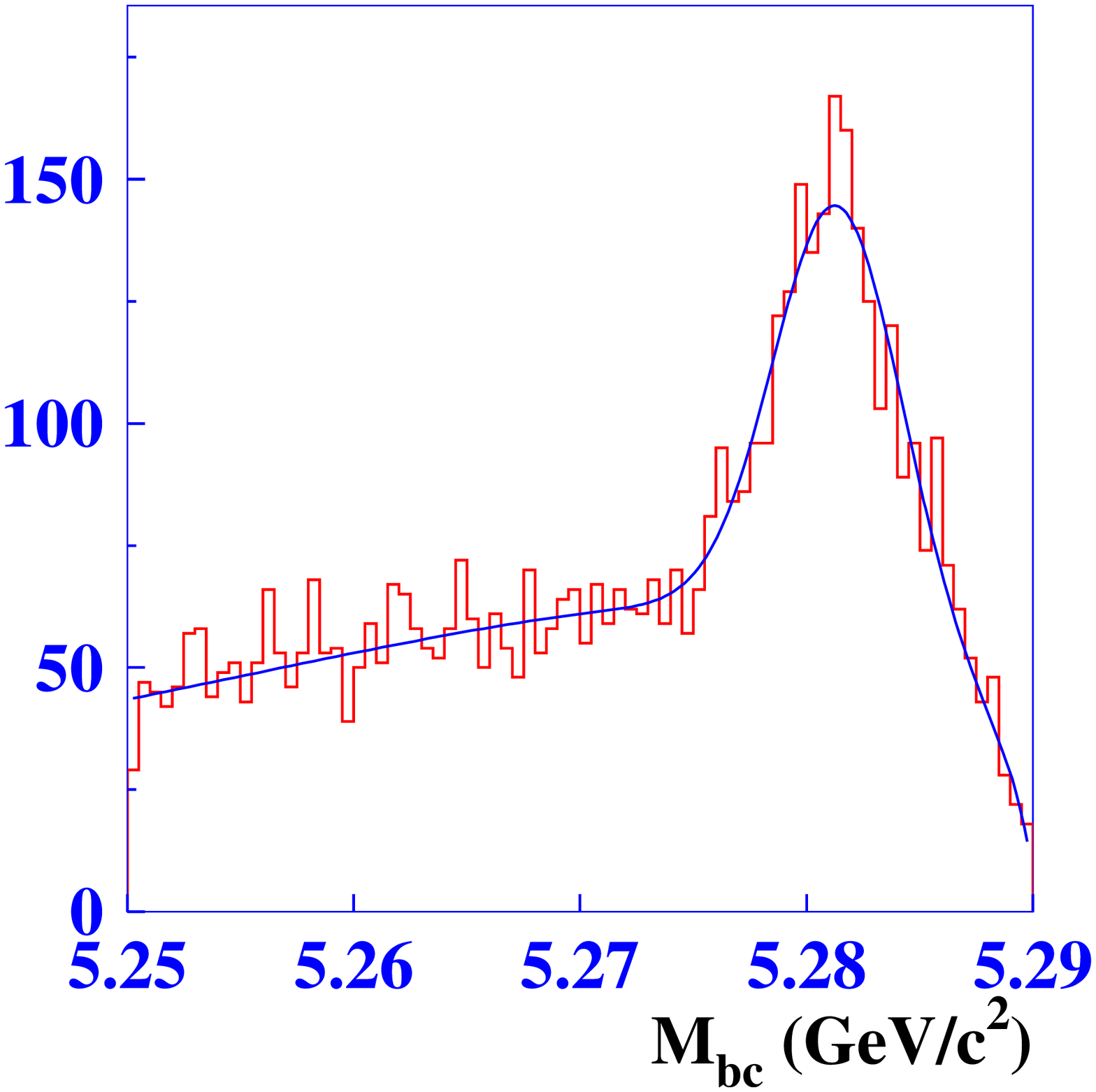}&
 \includegraphics[height=8 cm]{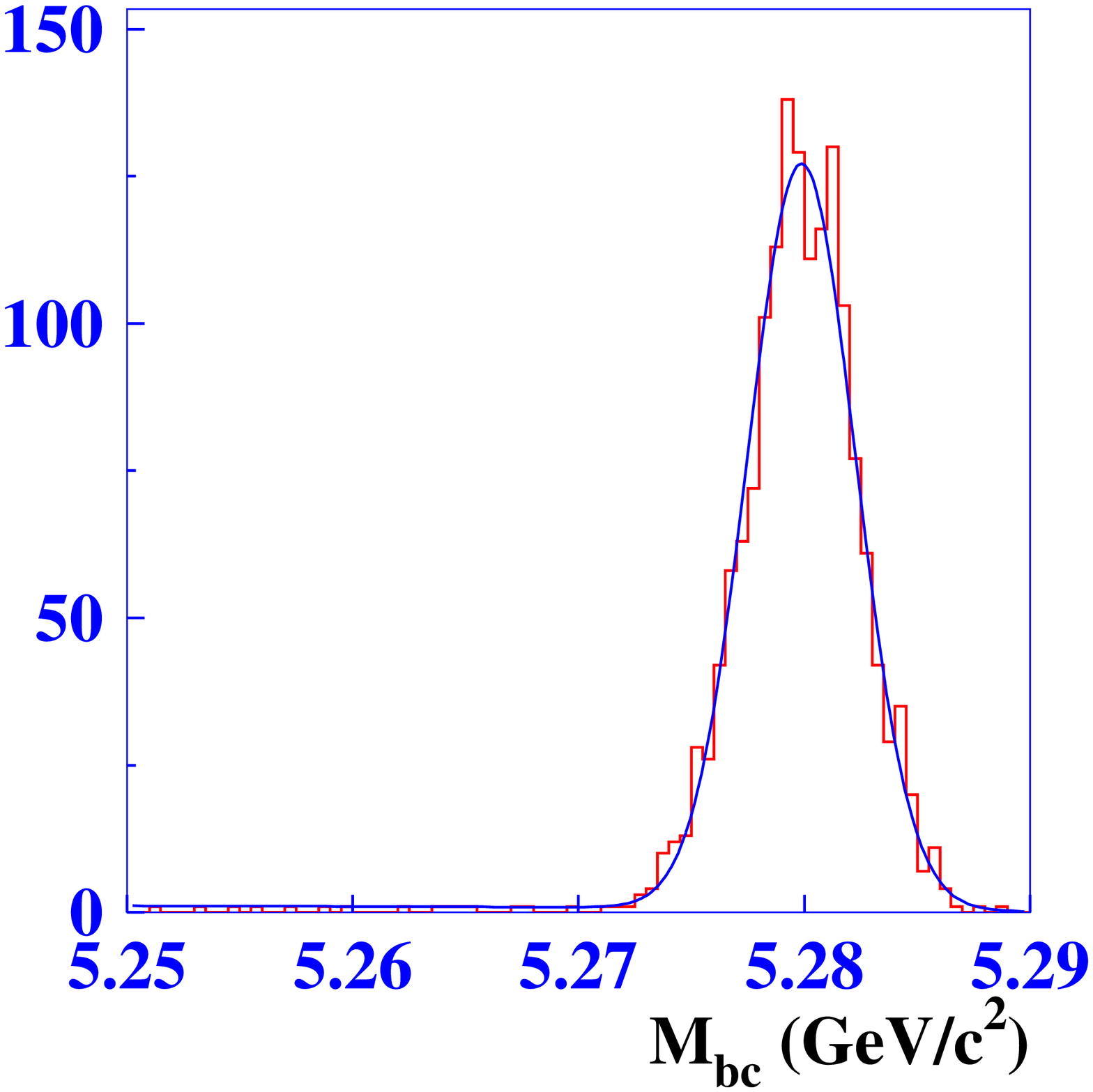}\\
 \includegraphics[height=8 cm]{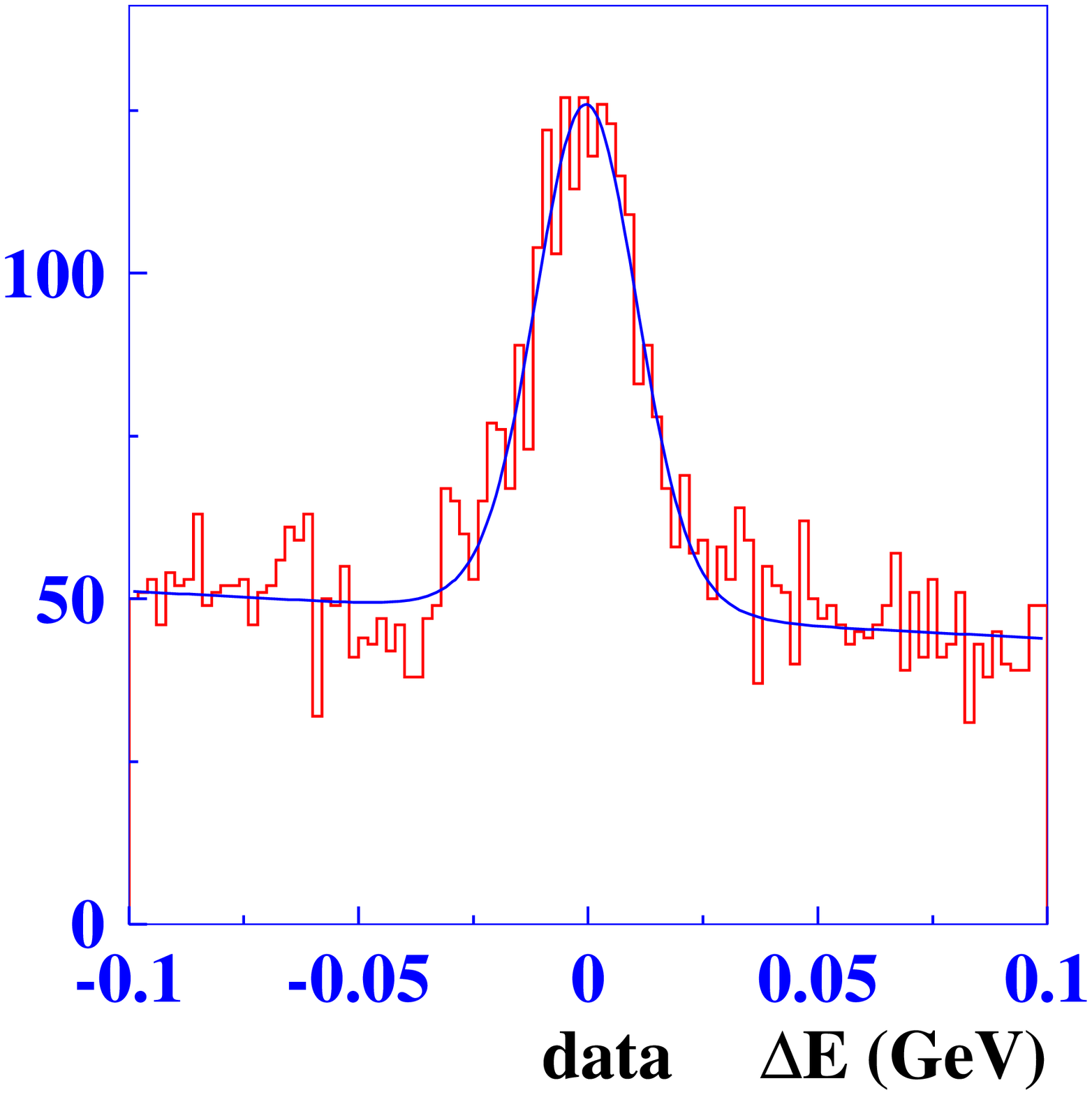}&
 \includegraphics[height=8 cm]{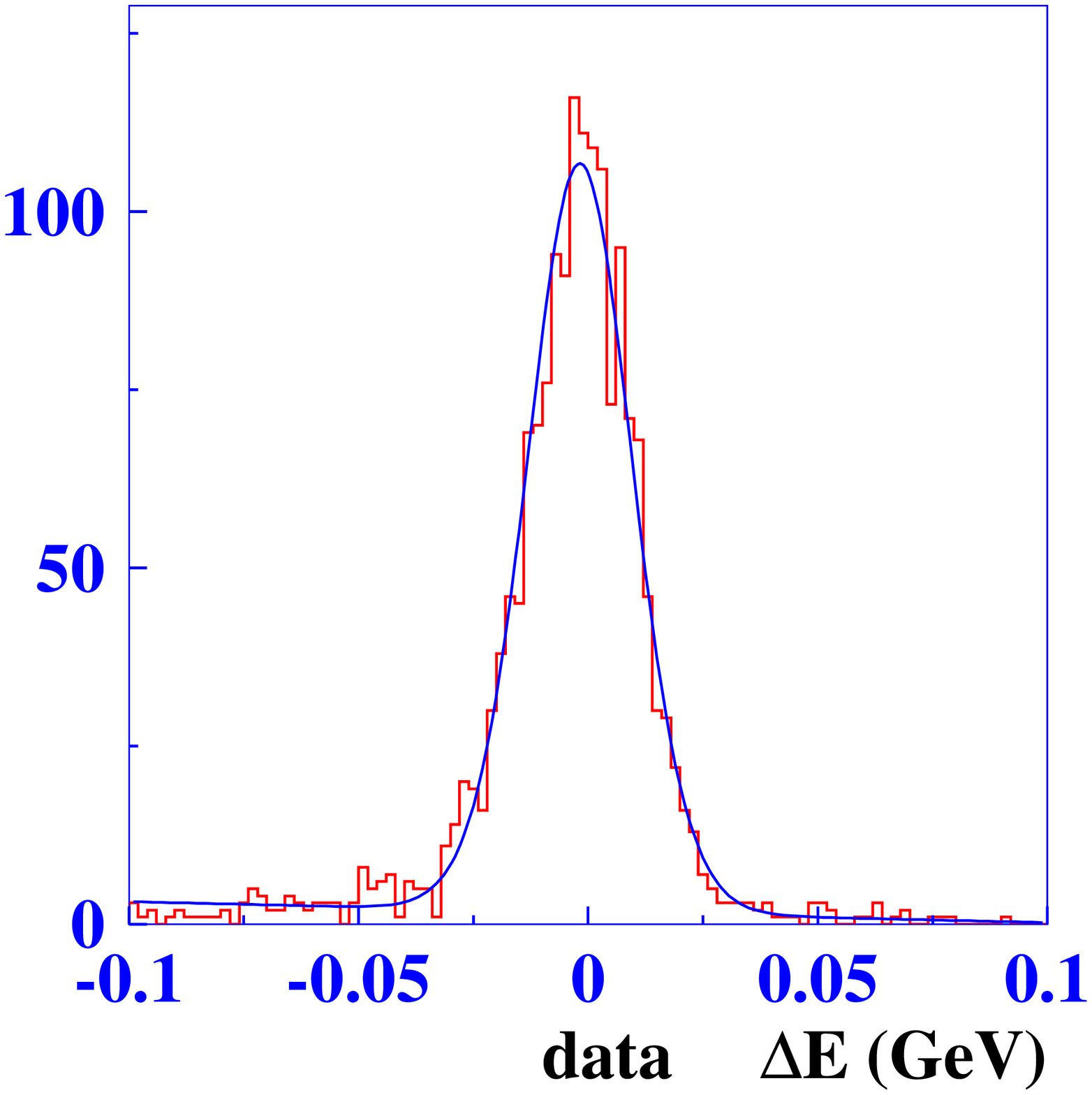}\\
\vspace*{-16 cm} & \\
{\bf\large \hspace*{-3cm} a)}&{\bf\large \hspace*{-3cm} b)}\\
\vspace*{7. cm} & \\
{\bf\large \hspace*{-3cm} c)}&{\bf\large \hspace*{-3cm} d)}\\
\vspace*{6.5 cm} & \\
\end{tabular}
\caption{$M_{\rm bc}$ and  $\Delta E$ distributions for
 $\bar{B}^{0}\to D^{0}\pi^{+}\pi^{-}$ events.
(a),(b) show $M_{\rm bc}$ distributions.
(c),(d) show $\Delta E$ distributions. 
(a),(c) ((b),(d)) distributions are plotted for sample (1)((2)), respectively.
 }
\label{f:dpmbde}
\end{figure}

 The $M_{\rm bc}$  and $\Delta E$ distributions
for $\bar{B}^{0}\to D^{0}\pi^{+}\pi^{-}$ events 
are shown in Fig.~\ref{f:dpmbde}. 
The distributions are plotted for events that satisfy 
the selection criteria for the other variable: i.e.,
$|\Delta E|<25$~MeV and $|M_{\rm bc}-M_B|<5$~MeV/$c^2$ for the 
$M_{\rm bc}$ and  the $\Delta E$ histograms, respectively. A clear signal 
is evident in both distributions.
The signal yield is obtained  by fitting the  $\Delta E$ 
distribution to the  sum of  two Gaussians
with the same mean for the signal and a linear
function for background. The widths and the relative normalization of the two Gaussians 
are fixed at values obtained from the MC simulation while 
the signal normalization  as well as the 
constant term and slope  of the background linear function
are treated as free parameters.

The signal yields  are $1128\pm51$ and $1521\pm40$
for samples (1) and (2), respectively.
A detection efficiency of $(16.8\pm 0.4)\%$ and $(18.2\pm 0.4)\%$
is determined from a MC
simulation that uses a Dalitz plot 
distribution that is 
 generated according to the model described 
in the next section.
Taking into account the branching fraction 
${\cal B}(D^0\to K^-\pi^+)=(3.80\pm0.09)\%$ and ${\cal B}(D^{*+}\to D^0\pi^+)=(67.7\pm0.5)\%$~\cite{PDG},   
we obtain the following value for the branching fractions: 
$$
{\cal B}(\bar{B}^0\to D^0\pi^+\pi^-) =(1.07\pm0.06\pm0.10)\times10^{-3},
$$
and
$$
{\cal B}(\bar{B}^0\to D^{*+}\pi^-)=(2.30\pm0.06\pm0.19)\times10^{-3},
$$
where the first error is statistical and second error is
systematic. The 
main contributions 
to the systematic error are listed in 
Table~\ref{t:sys}.
\begin{table}
\begin{center}
\begin{tabular}{l|l|l}
\hline
Source &\multicolumn{2}{|c}{$\sigma_{sys},~\%$}\\
\hline
& sample(1)&sample(2)\\
\hline
PID & $5\%$        & $5\%$ \\        
Background & $5\%$ & $1\%$ \\  
Tracking  & $4.4\%$   & $5.4\%$ \\
MC  & $3\%$ 	   & $3\%$ \\  
Br($D,D^*$)&$2.4\%$&$2.5\%$ \\  
\hline		   		       
Total& $9.2\%$ 	   & $8.1\%$ \\
\hline
\end{tabular}
\end{center}
\caption{The systematic uncertainties for $\bar{B}^0\to D^0\pi^+\pi^-$.}
\label{t:sys}
\end{table}
The uncertainty of the background shape 
was estimated by adding higher order polynomial terms to the approximating 
function.

The values of ${\cal B}(\bar{B}^0\to D^0\pi^+\pi^-)$ and ${\cal B}(\bar{B}^0\to D^{*+}\pi^-)$ 
are in agreement with a  previous result from Belle:
$
{\cal B}(\bar{B}^0\to D^0\pi^+\pi^-)=(8.0\pm1.6)\times10^{-4}
$~\cite{Asish} and a 
CLEO result: ${\cal B}(\bar{B}^0\to
D^{*+}\pi^-)=(2.81\pm0.25)\times10^{-4}$~\cite{CLEOd1} and are more precise.

\subsection{$B\to D\pi\pi$ Dalitz plot analysis}
For a three-body decay of a spin zero particle, two 
variables are required to describe the decay kinematics; 
we use the 
$D^0\pi^+$ and $\pi^+\pi^-$ invariant masses squared, $q^2$ and $q_1^2$,
respectively.

To analyze  the dynamics of $B\to D\pi\pi$ decays,  events 
with $\Delta E$ and $M_{\rm bc}$
within the signal region
$((\Delta E+\kappa(M_{\rm bc}-M_B))/\sigma_{\Delta E})^2+((M_{\rm
  bc}-M_B)/\sigma_{M_{\rm bc}})^2<4$ 
are selected. The parameters $\sigma_{\Delta E}=11
{\rm MeV}/c^2,~\sigma_{M_{\rm bc}}=2.7~{\rm MeV}/c^2,~\kappa=0.9$  have been
obtained from a fit to experimental data; the coefficient $\kappa$ takes into account
a correlation between  $M_{\rm bc}$ and $\Delta E$.  

\begin{figure}[h]
\begin{center}
\includegraphics[width=8 cm]{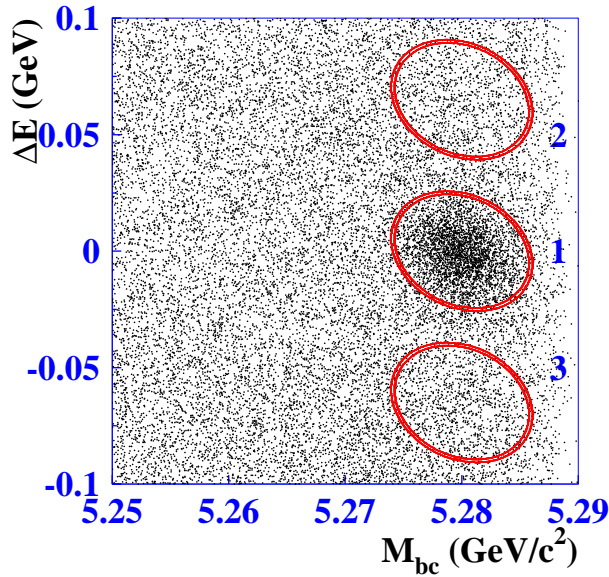}
\caption{The experimental event distribution in the $ (M_{\rm bc}-\Delta
  E))$ plot. The ellipses show a position of the signal (1) and sideband regions (2),(3).}
\label{f:mbc}
\end{center}
\end{figure}

To model  the contribution and shape of the background, we use
events from the $\Delta E$ sidebands, which are defined as:
$((\Delta E\pm65\,{\rm MeV}+\kappa(M_{\rm bc}-M_B))/\sigma_{\Delta E})^2+((M_{\rm
  bc}-M_B)/\sigma_{M_{\rm bc}})^2<4$. Figure~\ref{f:mbc} shows the 
signal and sidebands regions in the $ (M_{\rm bc}-\Delta E))$ plane.

\begin{figure}[h]
\begin{center}
\begin{tabular}{cc}
\includegraphics[width=8 cm]{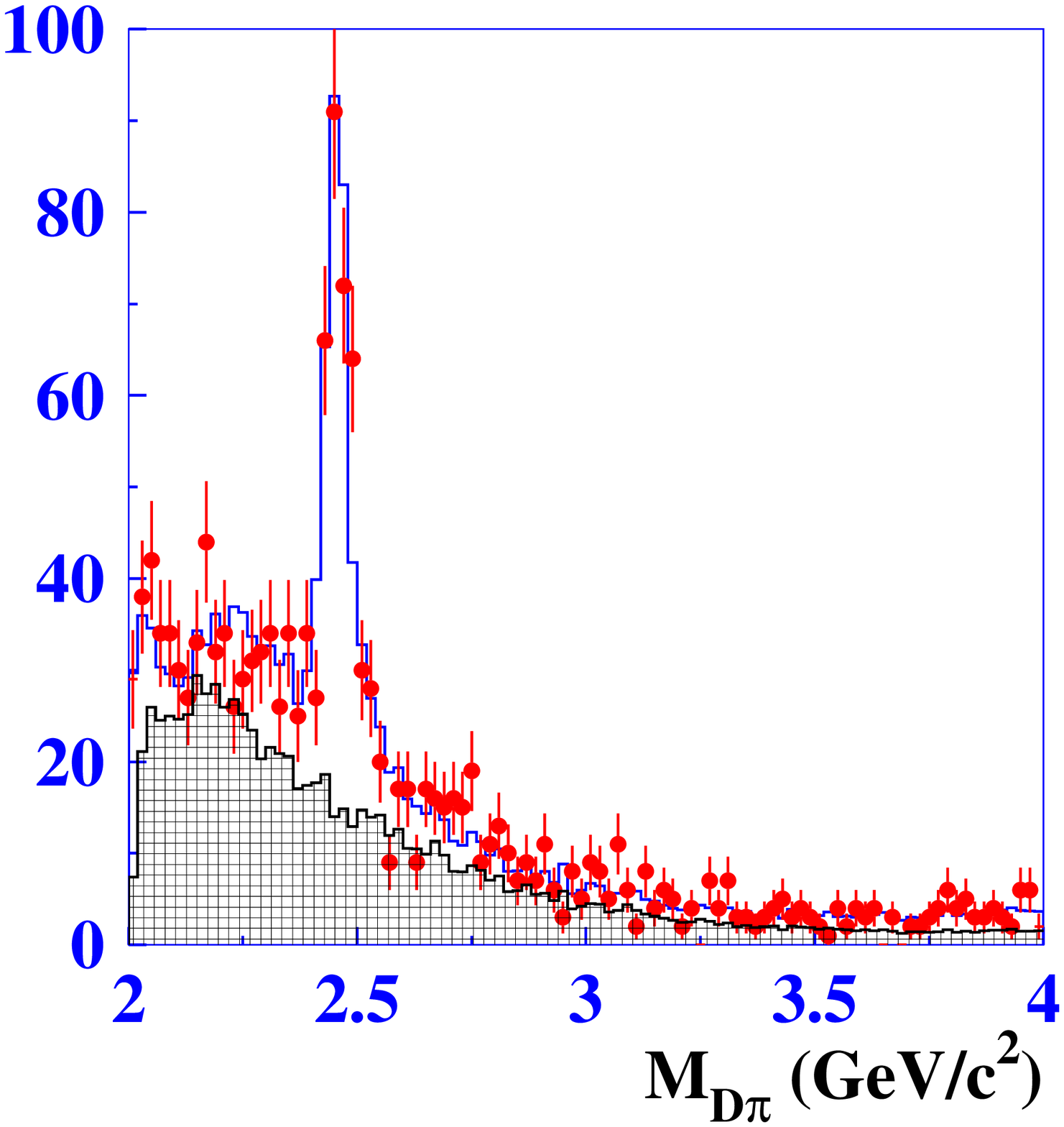}&
\includegraphics[width=8 cm]{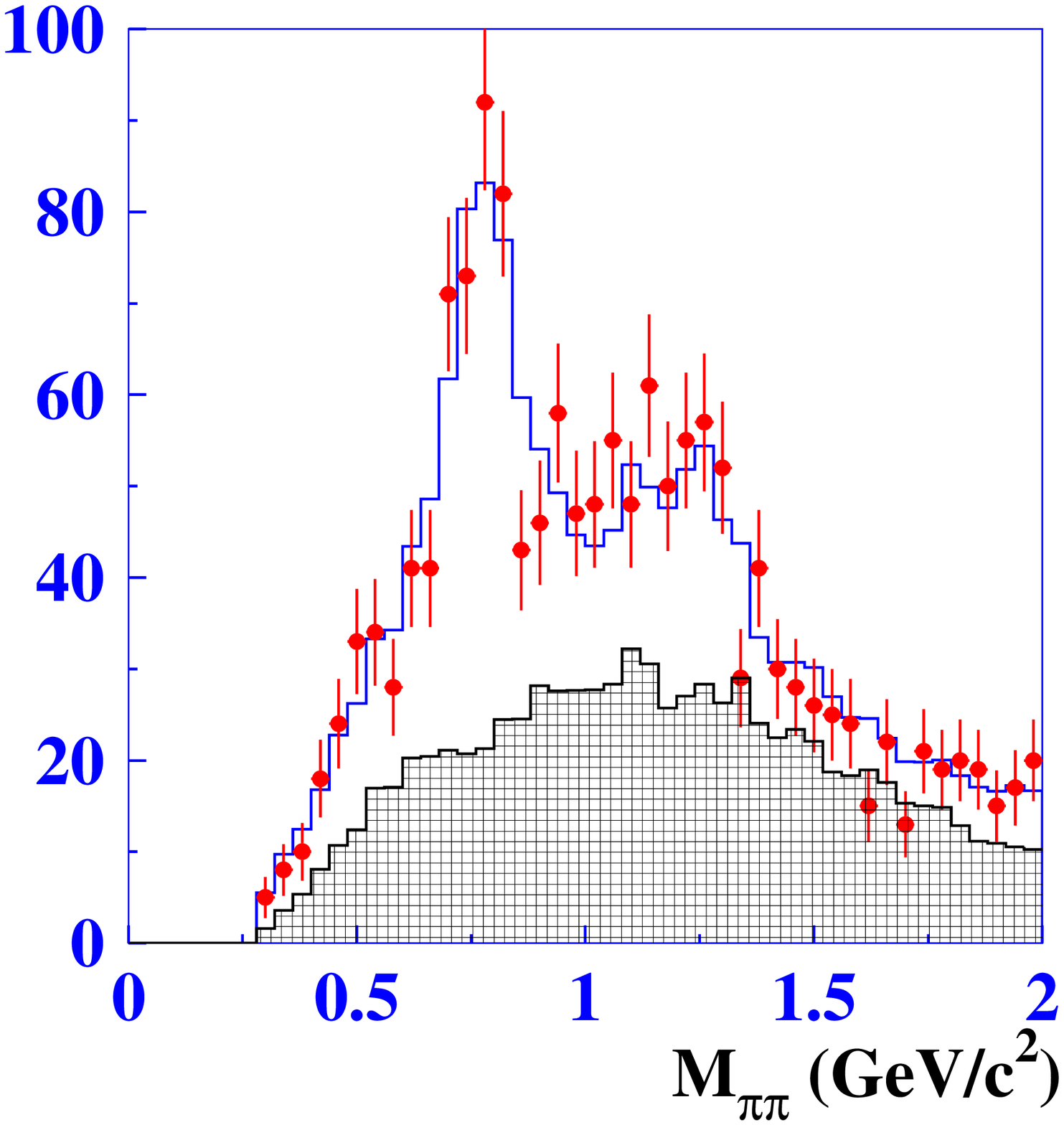}\\
\vspace*{-8 cm} & \\
{\bf\large \hspace*{-3cm} a)}&{\bf\large \hspace*{-3cm} b)}\\
\vspace*{6.5 cm} & \\
\end{tabular}
\caption{$D\pi$ - (a) and $\pi\pi$ -(b) mass distribution. The points
  with error bars
  correspond to the signal box events,  the hatched 
histogram --- to the background obtained from sidebands. The open histogram
shows  the fit function after efficiency correction.
}
\label{f:dpp_M}
\end{center}
\end{figure}

The $D\pi$ and $\pi\pi$ mass distributions for the signal and sideband events 
are shown in Fig.~\ref{f:dpp_M}. In the  $D\pi$ mass distribution we
can clearly see the narrow peak of the $D_2^*$.  The $\pi\pi$ distribution
has a signal of the $\rho$ meson as well as a 
structure at $1.2-1.3\,\rm GeV/c^2$ that can be due to $f_0(1370)$ or
$f_2(1270)$ contributions.

\begin{figure}[h]
\begin{center}
\begin{tabular}{cc}
\includegraphics[width=8 cm]{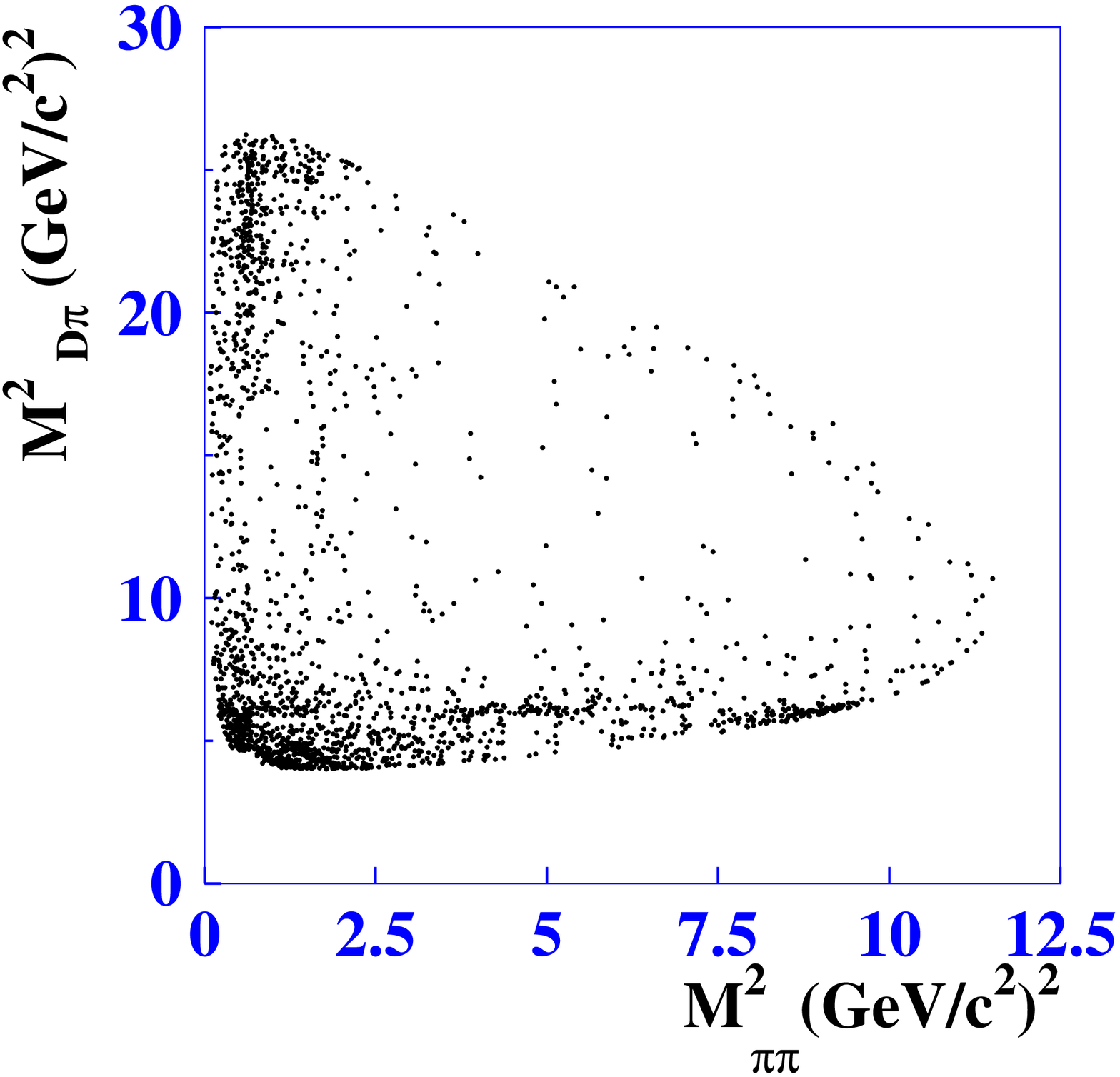}&
\includegraphics[width=8 cm]{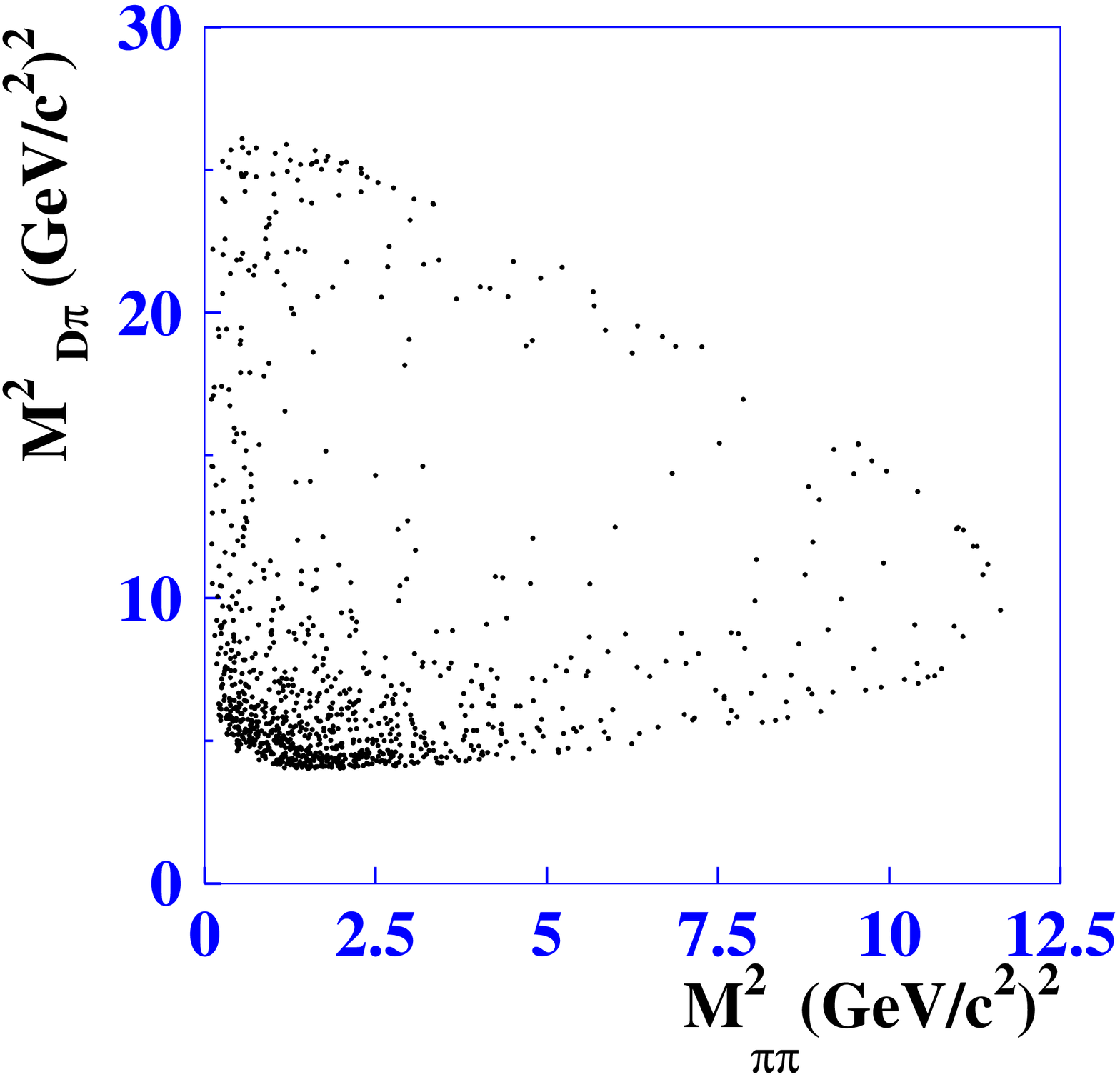}\\
\vspace*{-8 cm} & \\
{\bf\large \hspace*{-3cm} a)}&{\bf\large \hspace*{-3cm} b)}\\
\vspace*{6.5 cm} & \\
\end{tabular}
\caption{The Dalitz plot of a)signal events; b)sideband events.}
\label{f:dpp_DP}
\end{center}
\end{figure}

The distributions of events in the $M^2_{D\pi}$ versus $M^2_{\pi\pi}$
Dalitz plot for the signal and sideband regions
are shown 
in Fig.~\ref{f:dpp_DP}. 
The Dalitz plot boundary is determined by the
decay kinematics and the masses of the daughter particles. 
In order to have the same Dalitz plot boundary for events  
in both signal and sideband regions,
mass-constrained fits 
of 
$K\pi$ to $M_D$ and $D\pi\pi$ to 
$M_B$ are performed. 
The mass-constrained fits also reduce smearing from
detector resolution. 

To extract the amplitudes and phases of different intermediate states,
an   unbinned fit to the Dalitz plot is performed using the method
described in Ref.~\cite{mybelle}.
The event density function in the Dalitz plot is the sum of the signal  
and  background.

The background distribution and normalization are obtained from the  $\Delta E$  sideband
analysis.
Since the  $D\pi$ mass distributions for the upper  
and lower halves of the $\Delta E$  sideband have similar shapes, we
can expect similar background behavior for the signal and sideband regions.
The background Dalitz plot has neither a resonant structure
nor non-trivial helicity behavior and is combinatorial in its origin.
The background shape 
is obtained from an unbinned fit of 
the sideband distribution to a smooth two-dimensional function.
The number of background events in the signal region 
is scaled according to the relative areas of the signal and the 
sideband regions. 

There is no generally accepted way to exactly describe a three-body
amplitude. In this paper we represent the $D\pi\pi$ amplitude as
the sum of Breit-Wigner contributions for different intermediate
two-body states.  Such
an approach cannot be exact since it is neither analytic nor unitary and
does not take describe completely possible final state interactions.
 Nevertheless, 
the sum of Breit-Wigners describes the main features of the
amplitude behavior and allows one to find and distinguish the
contributions of various two-body intermediate states, their interference
and the effective parameters of these states. We used the same approach 
in the analysis of charged $B$ decays~\cite{mybelle}.

In the $D^0\pi^+\pi^-$ final state 
a  combination of the  $D^0$-meson and a pion can form a vector meson $D^{*+}$, a tensor meson 
$D^{*+}_2$ or a scalar state $D_0^{*+}$; the axial vector mesons $D_1^+$ and 
$D'^{+}_1$ cannot decay to two pseudoscalars because of angular 
momentum and parity conservation. The region of the $D^{0}\pi^+$
invariant mass
corresponding to the $D^{*+}$
is excluded from the fitting by requiring $|M_{D\pi}-M_{D^*}|>0.01~\rm
GeV/c^2$  but  in $B$ decay a
virtual $D^{*+}$ (referred to as $D^*_v$)  can be produced off-shell
with $\sqrt{q^2}$
larger than the $D^0\pi^+$ total mass and such a process will
contribute to the amplitude. 
Another virtual hadron that can be produced in this combination is $B^{*-}$ (referred to as $B^{*}_v$):
$B\to B^*_v\pi$ and $B^*_v\to D\pi$. 
For the
mass of ${B}^{*-}$ as well as the mass and width of $D^{*+}$, we used the PDG
values~\cite{PDG}; the widths of ${B}^{*-}$ are
calculated from the width of the $D^{*-}$ in the HQET approach.
In the $\pi\pi$ distribution we can see a $\rho$ meson signal and some
evidence  of resonances around $1300~\rm MeV$, which we describe with the 
hypothesis 
of 
$f_2(1270)$ and $f_0$(1370) mesons. As it is shown in 
Table~\ref{t:dpp1},
the hypothesis of $f_2$ has the best likelihood value. We also include the scalar resonance $f_0$(600) with free mass and width.

\begin{table}
\begin{tabular}{|c|c|c|c|c|c|c|c|c|}
\hline
&\multicolumn{8}{|c|}{$D^*_2,~D^*_v,~\rho,~f_0(600)$}\\
\hline
&\multicolumn{3}{|c|}{$D^*_0$,}&&\multicolumn{4}{|c|}{$f_2(1270)$}\\
\hline
&$f_2(1270)$&$f_0$(1370)& -- && $D_0$& $D_1$ & $D_2$& -- \\

\hline
$ -2\ln{\cal L}/{\cal L}_0$& 0 &49&80&&0&-5&16&31\\
\hline
\end{tabular}
\caption{Comparison of the models with different resonances.}
\label{t:dpp1}
\end{table}

The contributions from the intermediate states listed above 
are included in the signal-event density ($S(q^2,q^2_1)$)
parameterization 
as a coherent sum of the 
corresponding amplitudes
together with a possible constant amplitude ($a_{ps}$):
\begin{eqnarray}
S(q^2,q_1^2)&=&|a_2A^2(q^2,q_1^2)+a_0e^{i\phi_0}A^0
(q^2,q_1^2)+a_1e^{i\phi_1}A^1(q^2,q_1^2)\nonumber\\
&&+a_{\rho}e^{i\phi_{\rho}}A^{\rho}+a_{f_0}e^{i(\phi_{f_0}+\phi_{\rho})}A^{f_0}
+a_{f_2}e^{i(\phi_{f_2}+\phi_{\rho})}A^{f_2}\nonumber\\
&&+a_{B^*}e^{i\phi_B}A^1(q^2,q_1^2)+a_{ps}e^{i\phi_{ps}}|^2.
\end{eqnarray}
Each resonance is described by a relativistic Breit-Wigner with a
$q^2$ dependent width 
and angular dependence that corresponds to the spins of the
intermediate and final state particles following the approach described 
in~\cite{mybelle}.
We take into account hadronic transition form factors. 
The Blatt-Weisskopf parameterization~\cite{blat} with
a hadron scale $r$=1.6~$\rm(GeV/c)^{-1}$ is used.
For the virtual mesons $D^*_v$ and $B^*_v$ that are produced beyond the
peak region,
another form factor parameterization is used: 
\begin{equation}
\label{eqf1}
F_{AB}(\mathbf{p})=e^{-r(\mathbf{p-p_0})};
\end{equation}
this provides stronger suppression of the Breit-Wigner far from the
resonance region. 

The detector resolution for the invariant mass  of 
the $D\pi $($\pi\pi $) combination is about $2.5$ (3.5)~MeV  which is
much smaller than the narrowest peak width  30--40~MeV.

The masses and widths of the $\pi\pi$ resonances (except $f_0(600)$) are fixed at the PDG~\cite{PDG}
values.   The mass and width of the broad resonance in $(D\pi)$:   
$M_{D_0^{*+}}=2308\,\rm MeV/c^2,~\Gamma^0_{D_0^{*+}} =276\,{\rm MeV}/c^2$ have been taken from our
measurement for the $D^{**0}$~\cite{mybelle}.

The masses and widths of the $D_2^+$ and  the $f_0$ as well as the relative amplitudes and 
phases are free parameters of the 
fit.   The parameters of the the $f_0$ are quite uncertain and it can also be regarded as 
a nonresonant     S-wave structure.

Table~\ref{t:dpp} gives the results of the fit for different models.
The contributions of different states are characterized by the
branching 
fractions, which are
defined as:
\begin{equation}
\label{e:brf}
Br_i= \frac{a_i^2 \int |A_i(Q)|^2 dQ}{\int |\sum_i a_ie^{i\phi_i} A_i(Q)|^2 dQ},
\end{equation} 
where $A_i(Q)$ is the corresponding amplitude, $a_i$ and $\phi_i$ are
the amplitude coefficients and phases obtained from the fit. The
integration 
is performed over all the available phase 
space characterized by the multidimensional vector $Q$ (for decay to 3
spinless particles $dQ\equiv dq^2dq_1^2$), and $i$ is one of the intermediate states: $D^{*}_2,~D^{*}_0,~\rho,~f_2,~f_0,~D^*_v,~B^*_v$ or 
the constant term $a_{ps}$. 
\begin{table}
 \begin{tabular}{|c|c|c|c|c|c|}
 \hline
 & 1 & 2
 &3       & 4      & 5       \\
 \hline
 & $D^*_2,~D^*_0,~D^*_v,$ &
 $D^*_2,~D^*_0,$
 &$D^*_2,~D^*_0,~D^*_v,$       & $D^*_2,~D^*_0,~D^*_v,$       & $D^*_2,~D^*_v,$       \\
 & $\rho,~f_2,~f_0(600)$ &
 $\rho,~f_2,~f_0(600)$
 &$\rho,~f_2,~f_0(600),~B^*_v$       & $\rho,~f_2,~f_0(600)+ps$       & $\rho,~f_2,~f_0(600)$       \\
 \hline
  $ -2\ln{\cal L}/{\cal L}_0$ &   0    &   66.9    &   -9.0    &  -7.0     &  37.2    \\
 \hline
$Br_{D_2^*}(10^{-4})$    &   3.08$\pm$ 0.22&   3.23$\pm$ 0.22&   3.19$\pm$ 0.26&   3.07$\pm$ 0.22&   3.33$\pm$ 0.23\\
 \hline
$\phi_{D_0^*}$           &  -1.82$\pm$ 0.24&  -1.52$\pm$ 0.24&
 -2.15$\pm$ 0.28&  -1.81$\pm$ 0.20&   -- \\
$Br_{D_0^*}(10^{-4})$    &   0.60$\pm$ 0.17&   0.47$\pm$ 0.15&
 0.52$\pm$ 0.18&   0.62$\pm$ 0.15&   -- \\
 \hline
$\phi_{D^*_v}$           &  -1.57$\pm$ 0.25&   --&  -2.18$\pm$ 0.29&  -1.54$\pm$ 0.22&  -1.90$\pm$ 0.30\\
$Br_{D^*_v}(10^{-4})$    &   0.70$\pm$ 0.14&   --&   0.71$\pm$ 0.14&   0.71$\pm$ 0.12&   0.62$\pm$ 0.13\\
 \hline
$\phi_{\rho}$            &   1.83$\pm$ 0.24&   2.09$\pm$ 0.23&   1.23$\pm$ 0.22&   2.07$\pm$ 0.24&   1.90$\pm$ 0.24\\
$Br_{\rho}(10^{-4})$     &   2.91$\pm$ 0.28&   2.82$\pm$ 0.30&   2.50$\pm$ 0.33&   2.55$\pm$ 0.28&   2.97$\pm$ 0.31\\
 \hline
$\phi_{f_2}$             &   2.81$\pm$ 0.20&   3.05$\pm$ 0.19&   2.74$\pm$ 0.19&   2.74$\pm$ 0.19&   2.89$\pm$ 0.20\\
$Br_{f_2}(10^{-4})$      &   1.10$\pm$ 0.19&   1.31$\pm$ 0.20&   1.18$\pm$ 0.20&   1.12$\pm$ 0.17&   1.24$\pm$ 0.20\\
 \hline
$\phi_{f_0}$             &   0.34$\pm$ 0.18&   0.60$\pm$ 0.17&   0.34$\pm$ 0.17&   0.23$\pm$ 0.17&   0.50$\pm$ 0.20\\
$Br_{f_0}(10^{-4})$      &   1.75$\pm$ 0.26&   1.93$\pm$ 0.31&   2.28$\pm$ 0.31&   1.81$\pm$ 0.26&   1.62$\pm$ 0.28\\
 \hline
$\phi_{B^*_v}$           &   --&   --&   0.00$\pm$ 0.19&   --&   --\\
$Br_{B^*_v}(10^{-4})$    &   --&   --&   0.44$\pm$ 0.25&   --&   --\\
 \hline
$\phi_{ps}$              &   --&   --&   --&   1.00$\pm$ 0.21&   --\\
$Br_{ps}(10^{-4})$       &   --&   --&   --&   0.06$\pm$ 0.05&   --\\
 \hline
$M_{f_0},(GeV/c^2)$      &  0.658$\pm$0.062&  0.681$\pm$0.048&  0.649$\pm$0.056&  0.633$\pm$0.052&  0.676$\pm$0.057\\
$\Gamma_{f_0},(GeV/c^2)$ &   0.94$\pm$ 0.22&   0.72$\pm$ 0.14&
 1.00$\pm$ 0.20&   0.88$\pm$ 0.19&   0.78$\pm$ 0.17\\
\hline
\hline
 \end{tabular}
\caption{The fit results for different sets of amplitudes.}
\label{t:dpp}
\end{table}

In addition to the main minimum where the likelihood value is maximal  ${\cal L}_0$,
there are several local minima with smaller likelihood ${\cal L}$.
The value of $(-2\ln{\cal L})$ differs from $(-2\ln{\cal L}_0)$
 by  1.9-25. To search for local
minima we perform 100 minimizations starting from different points, randomly
distributed in the space of the minimized parameters. 
Local minima appear mainly as a result of  different phases between the
$D\pi$ and $\pi\pi$ structures. We consider the spread of the
branching values as a model error. The central values are the
parameters obtained for the main minimum.  The spread of the 
relative  phases is
rather large.   In some cases, the phases `''flip'' by close to $\pi$ radians,
so that the extraction of relative phases unreliable.

The values of the $D_2^{*+}$ resonance mass and width obtained from the fit are:
$$
M_{D^{*+}_2}=(2459.5\pm2.3\pm0.7^{+4.9}_{-0.5}) {\rm MeV}/c^2,~~\Gamma_{D^{*+}_2}=(48.9\pm5.4\pm4.2\pm1.9){\rm MeV}.
$$
These parameters are consistent with the measurements from the FOCUS experiment: 
$
M_{D^{*0}_2}=(2467.6\pm1.5\pm0.8)\, {\rm MeV}/c^2,~\Gamma_{D^{*}_2}=34.1\pm6.5\pm4.2)\, \rm{MeV}
$~\cite{FOC}.

The product of the branching ratios for $D^*_2$ production obtained in the
analysis when the amplitudes are included  from above is:
$$
{\cal B}(\bar{B}^0\to D^{*+}_2\pi^-)\times B(D_2^{*+}\to D^{0}\pi^+)=(3.08\pm0.33\pm0.09^{+0.15}_{-0.02})\times10^{-4},
$$
where the indicated  errors are the statistical, systematic and model error.

The broad resonance branching fraction assuming $0^+$ quantum numbers is:
$$
{\cal B}(\bar{B}^0\to D^{*+}_0\pi^-)\times B(D_0^{*+}\to D^{0}\pi^+)=(0.60\pm0.17\pm0.16^{+0.13}_{-0.31})\times10^{-4}.
$$

\begin{figure}[h]
\begin{center}
\begin{tabular}[c]{cc}
\includegraphics[height=8 cm]{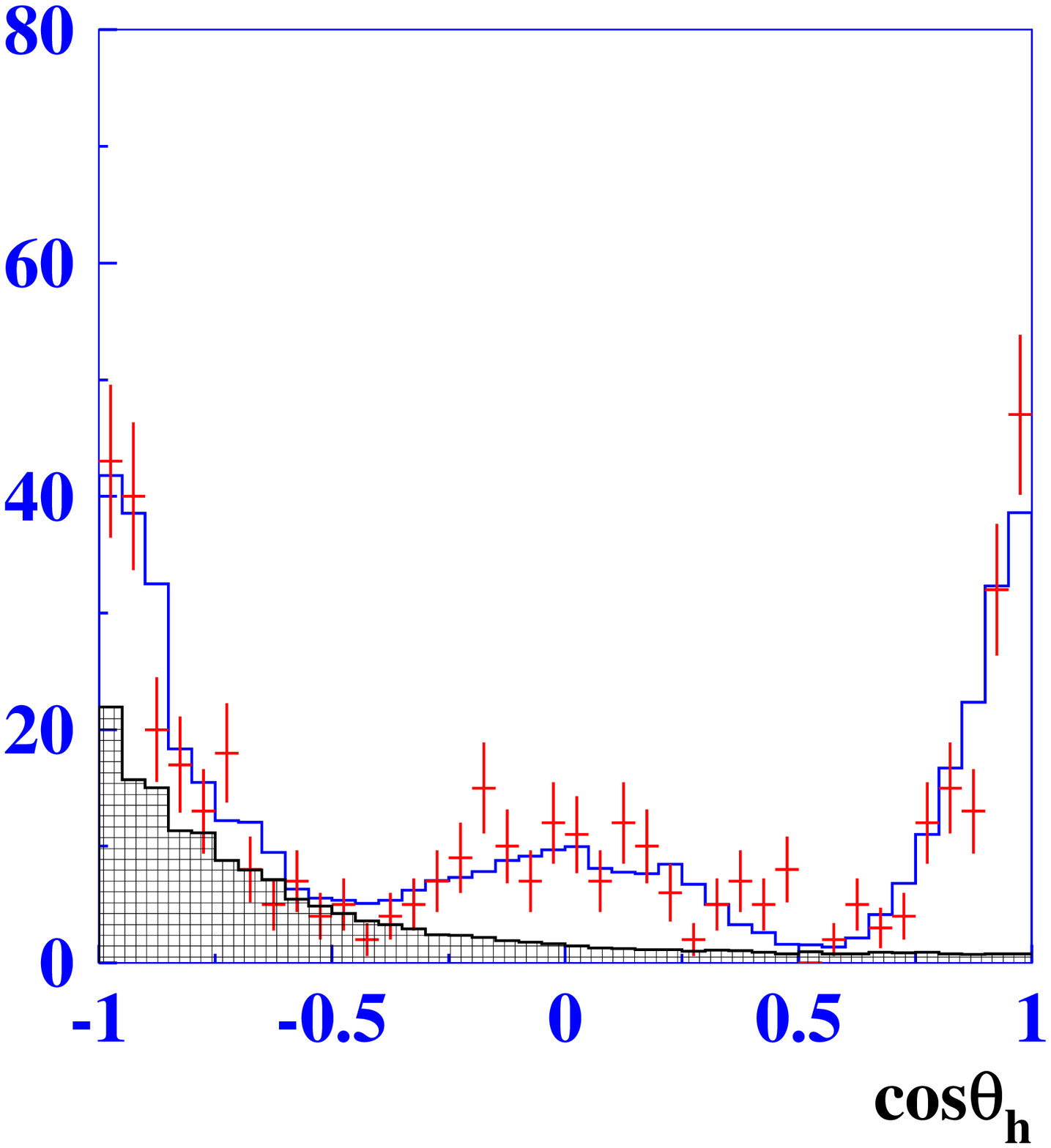}&
\includegraphics[height=8 cm]{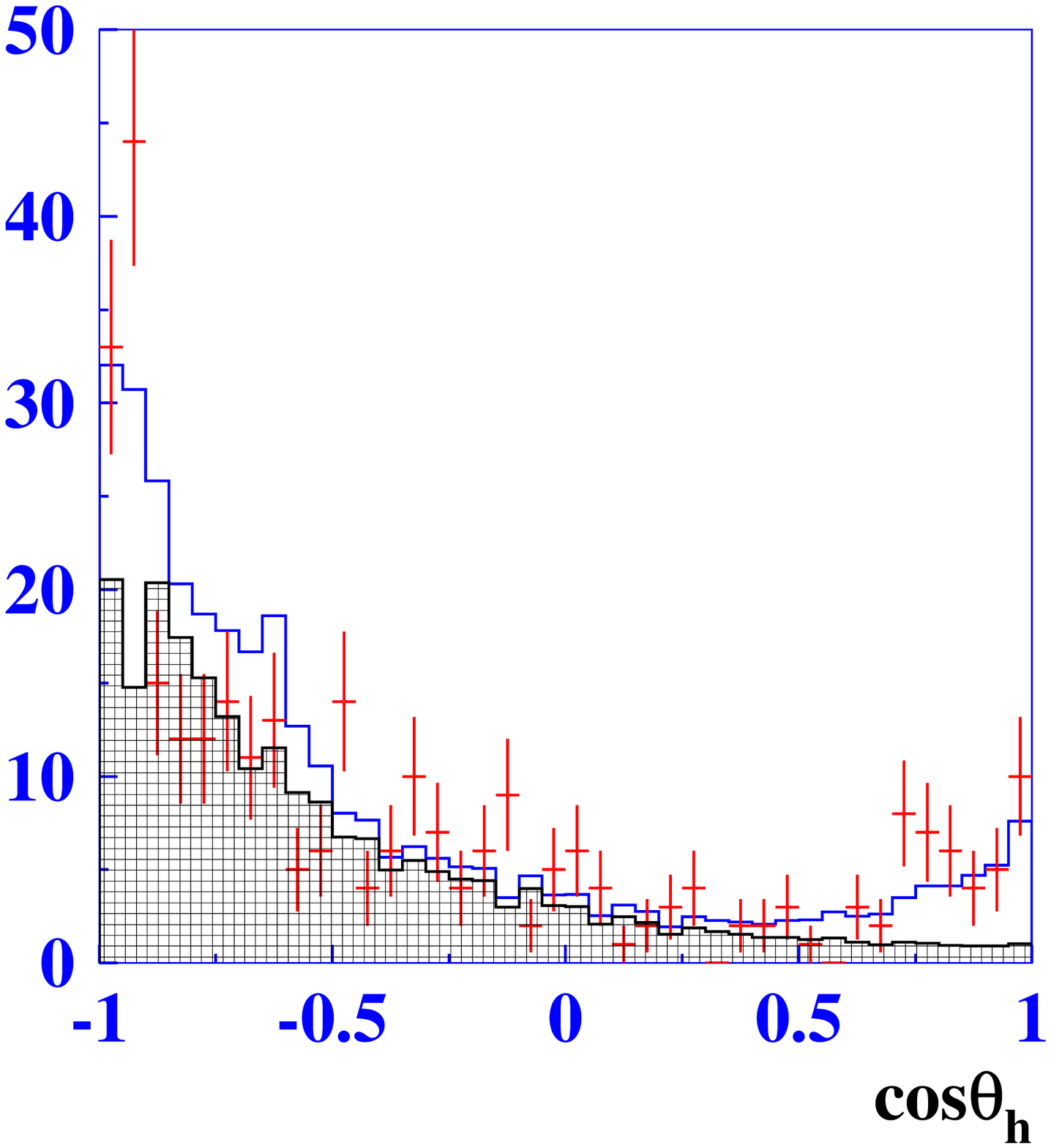}\\
\vspace*{-8 cm} & \\
{\bf\large \hspace*{-3cm} a)}&{\bf\large \hspace*{-3cm} b)}\\
\vspace*{6.5 cm} & \\
\end{tabular}
\end{center}
\caption{Helicity of the $D\pi$ distribution for experimental events
  (points) 
and for  MC simulation (histogram). The hatched distribution shows 
the background from the $\Delta E$ sideband region with 
a proper normalization. (a) corresponds to the $D_2$ region
$|M_{D\pi}-2.46|<0.1\rm\,GeV/c^2$; (b) - $D_0$ region $|M_{D\pi}-2.30|<0.1\rm\,GeV/c^2$.
} 
\label{f:dpp_hel}
\end{figure}

The helicity of the $D\pi$
distribution for different regions of $q^2$  is shown in
Fig.~\ref{f:dpp_hel} together with the efficiency corrected fitting function.
The histogram 
in the region of the $D^*_2$ meson clearly indicates a D-wave. The 
distributions in the other  region 
show  reasonable agreement of the fitting function 
and the data but the present statistics does not allow confirmation of the 
quantum numbers of the resonance.  Table~\ref{t:dpp1} shows that
the likelihood changed significantly if we exclude this resonance but
differs only slightly if the broad resonance is replaced with a vector or a
tensor. 
We thus set an upper limit for the 
branching fraction of the scalar $D_0^+\pi^-$ production:
$$
{\cal B}(\bar{B}^0\to D^{*+}_0\pi^-)\times B(D_0^{*+}\to D^{0}\pi^+)
<1.2\times10^{-4} \rm~at~90\,\%~C.L..
$$

The uncertainty of the background is one of the main sources of the systematic errors.
It is estimated by comparing the fit results for 
the case when the  background shape is taken 
separately from the lower or upper 
sideband in the $\Delta E$ distribution. The fit is also performed 
with more 
restrictive and loose cuts on $\Delta E$, $M_{\rm bc}$ and $\Delta M_D$
that changes the signal-to-background ratio by more than a factor of 2.
The obtained results are consistent with each other. 
The maximum difference is taken as an additional estimate 
of the systematic uncertainty.   
For the branching fractions, the systematic errors also include 
uncertainties on  track reconstruction and PID efficiency,
as well as the error in the $D^0\to K^-\pi^+$ absolute 
branching fraction.

The model uncertainties are estimated by comparing fit results 
for the case of different models
and for the values of  the parameter $r$ of the  transition form factor 
that range from 0 to 3 (GeV/$c$)$^{-1}$. 

\begin{figure}[h]
\begin{center}
\begin{tabular}[c]{ccc}
\includegraphics[height=5 cm]{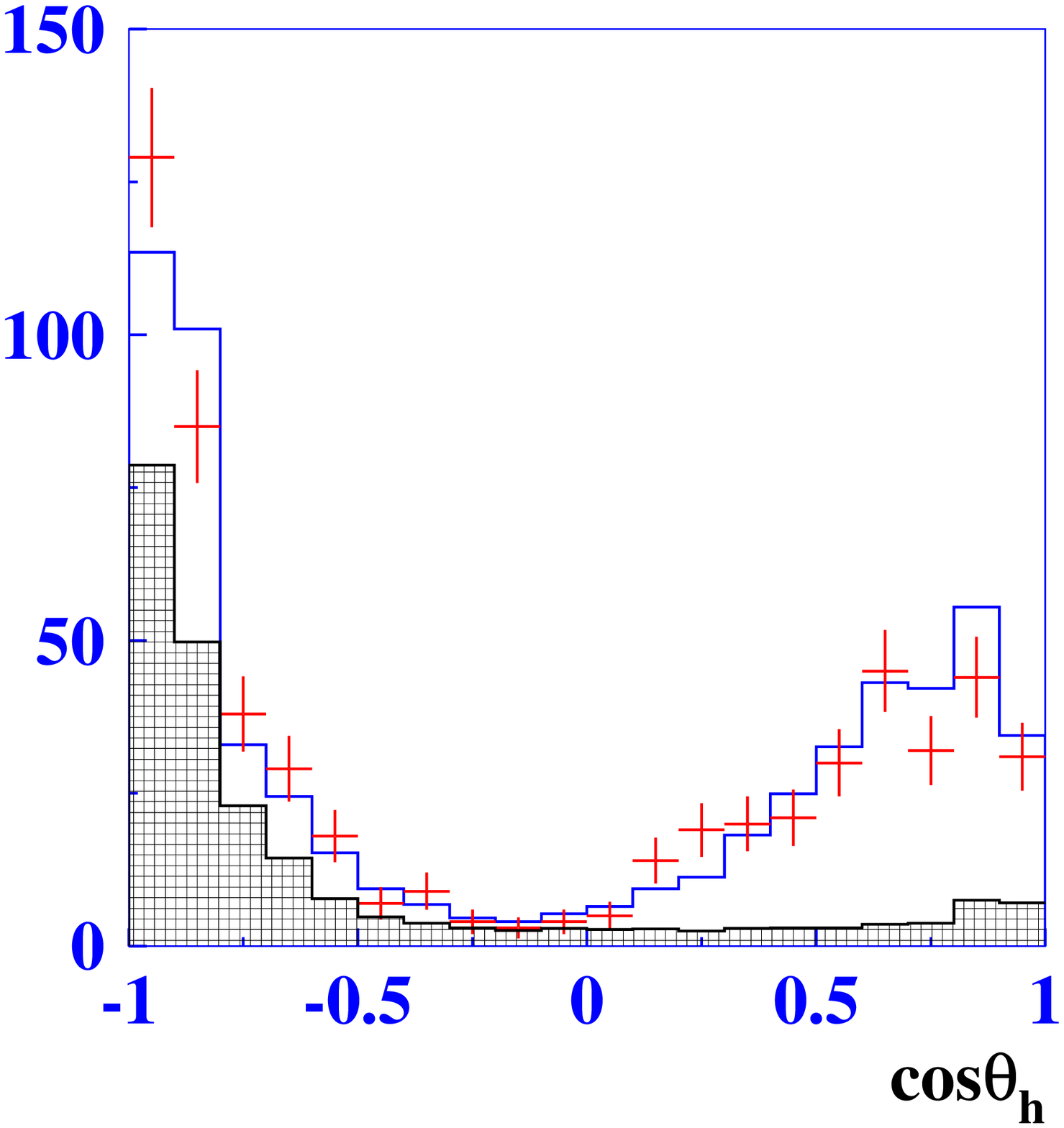}&
\includegraphics[height=5 cm]{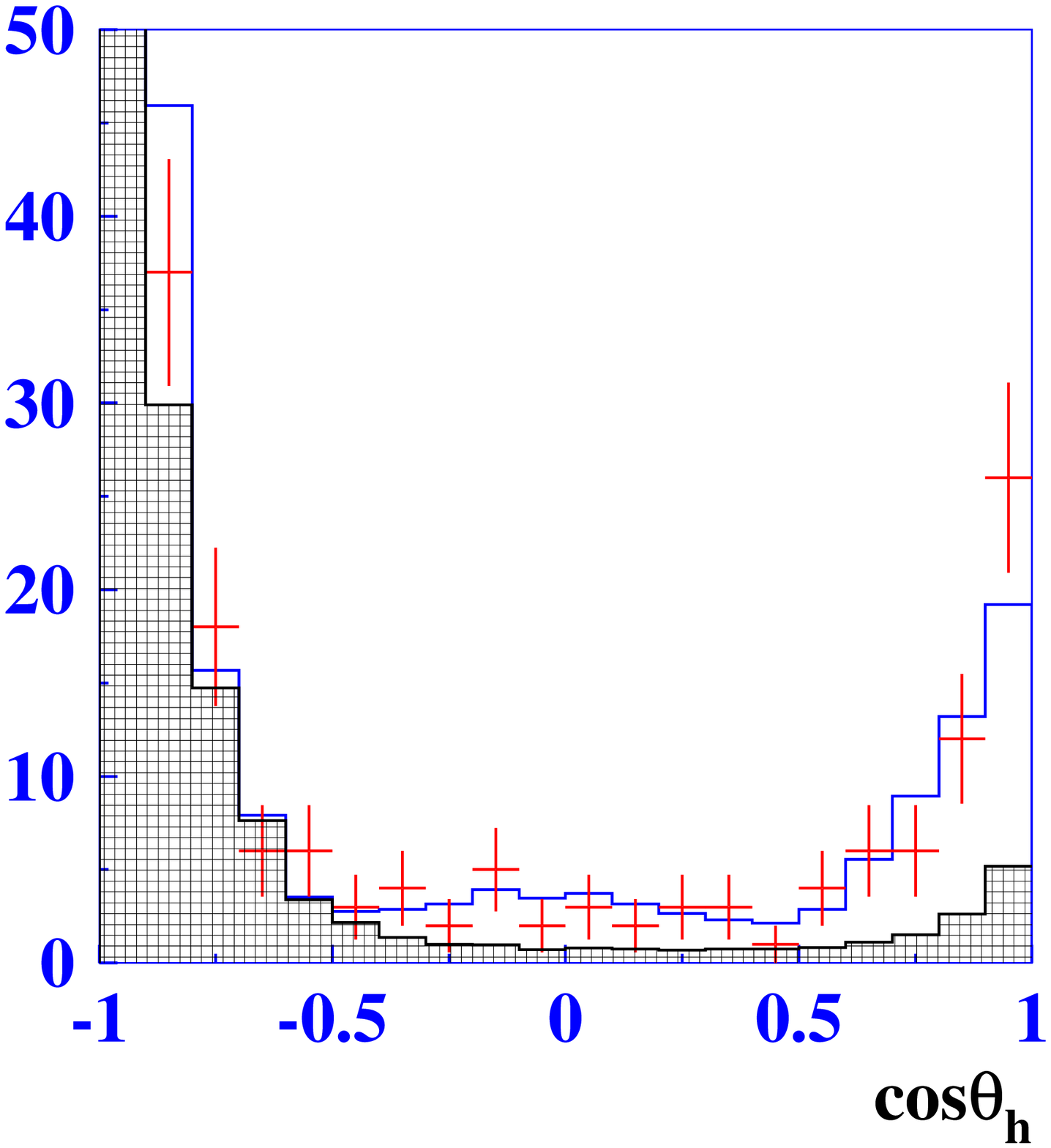}&
\includegraphics[height=5 cm]{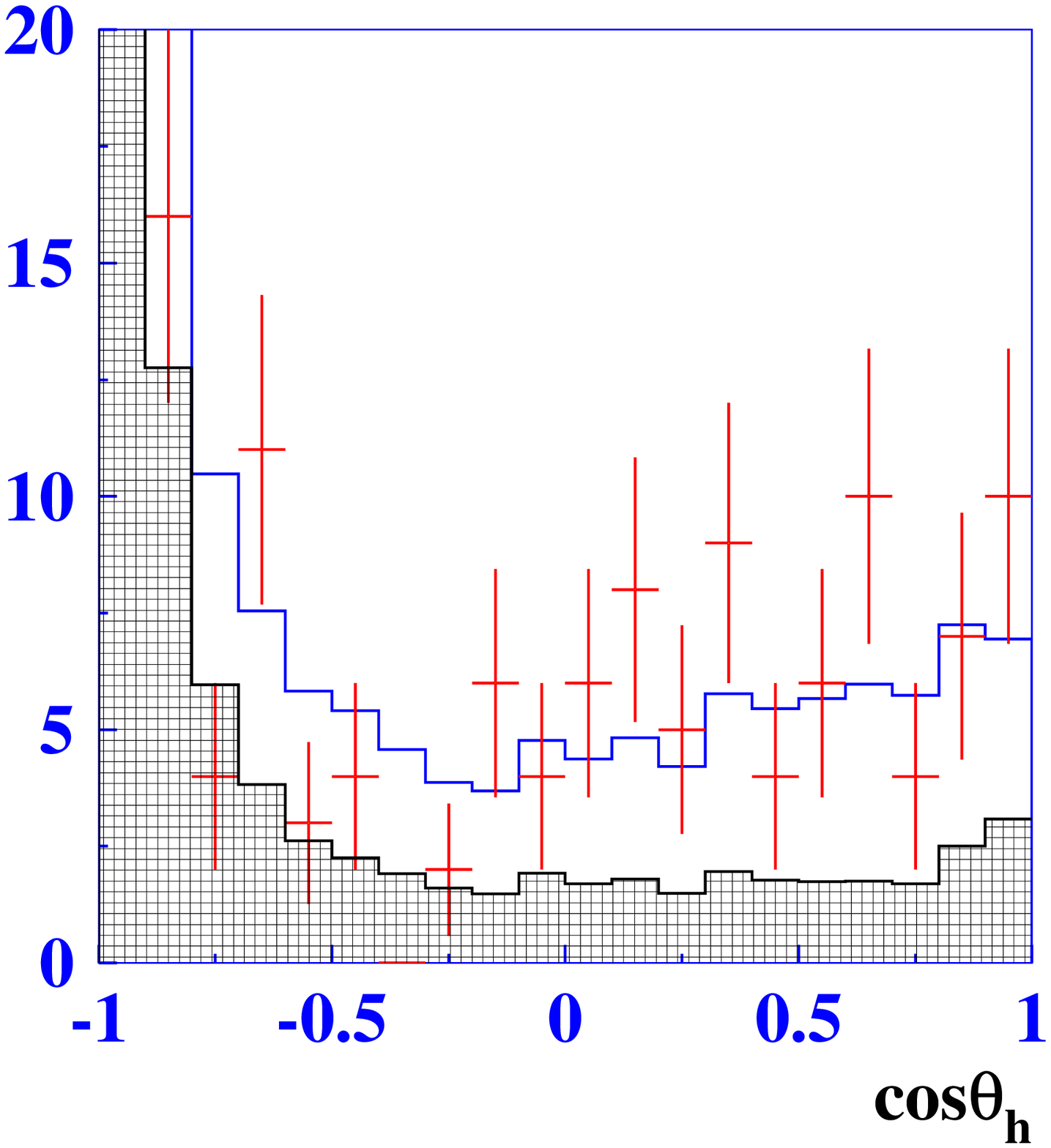}\\
\vspace*{-5 cm} & \\
{\bf\large \hspace*{-1cm} a)}&{\bf\large \hspace*{-1cm} b)}&{\bf\large \hspace*{-1cm} c)}\\
\vspace*{3 cm} & \\
\end{tabular}
\end{center}
\caption{Helicity distribution for experimental events (points) and for 
 MC simulation (histogram). The hatched distribution shows 
the background distribution from the $\Delta E$ sideband region with 
a proper normalization. (a) corresponds to the $\rho$ region $|M_{\pi\pi}-0.78|<0.2
\,$GeV/$c^2$; (b) -- $f_2$ region $|M_{\pi\pi}-1.20|<0.1\,$GeV/$c^2$; (c) -- $f_0$ region $M_{\pi\pi}<0.60\,$GeV/$c^2$.}
\label{f:pp_hel}
\end{figure}

The helicities of the $\pi\pi$ system in the $M_{\pi\pi}$ range
of $\rho,~f_2$ and below the $\rho$ where the broad resonance dominates
are shown in Fig.~\ref{f:pp_hel}.
For the positive helicity range where the $D\pi$ contribution is
suppressed, a clear $P$-wave structure for $\rho$ and 
$D$-wave for $f_2$ is observed.
The branching ratios for the $f_0(600)$ is 
$
{\cal B}(\bar{B}^0\to f_0 D^0){\cal B}(f_0\to \pi^+\pi^-)=
(1.75\pm0.26\pm0.35^{+0.55}_{-0.18})\times10^{-4}.
$ This process can also have a contribution from   some nonresonance background. 
The branching ratios for the $\rho$ and the $f_2$ are as follows: 
$$
{\cal B}(\bar{B}^0\to f_2 D^0){\cal B}(f_2\to \pi^+\pi^-)=
(1.10\pm0.19\pm0.21^{+0.18}_{-0.01})\times10^{-4}.
$$
Taking into account the PDG value of the branching fraction ${\cal B}(f_2\to \pi\pi)=$
$0.848^{+0.025}_{-0.013}$ and Clebsch-Gordan coefficients,we obtain:
$$
{\cal B}(\bar{B}^0\to f_2 D^0)=
(1.95\pm0.34\pm0.38^{+0.32}_{-0.02})\times10^{-4},
$$
$$
{\cal B}(\bar{B}^0\to \rho^0 D^0)=
(2.91\pm0.28\pm{0.33}^{+0.08}_{-0.54})\times10^{-4}.
$$

\section{$\bar{B}^{0}\to D^{*0}\pi^{+}\pi^{-}$ analysis.}

\begin{figure}[h]
\begin{center}
\begin{tabular}{cc}
\includegraphics[height=8 cm]{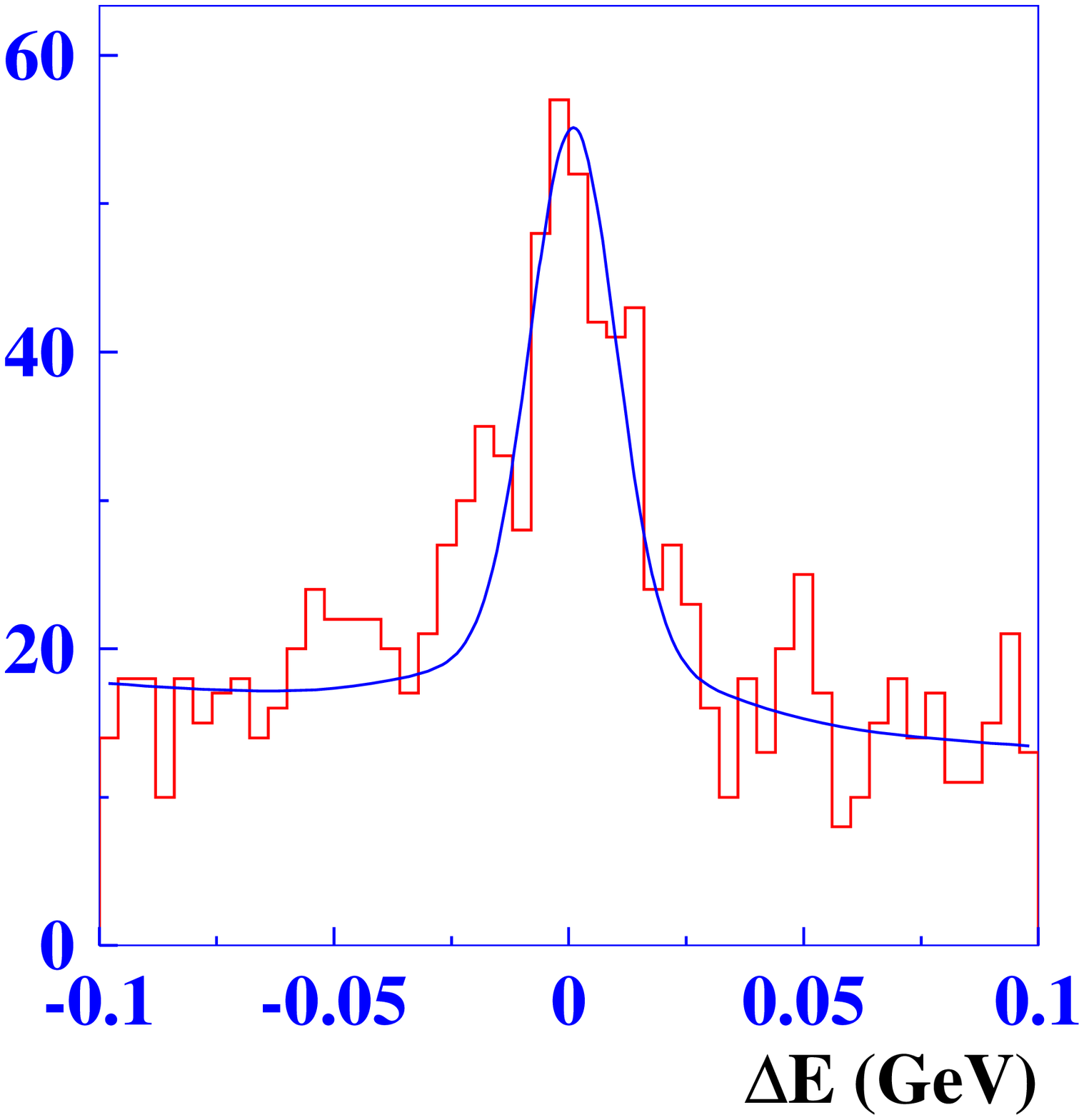}&
\includegraphics[height=8 cm]{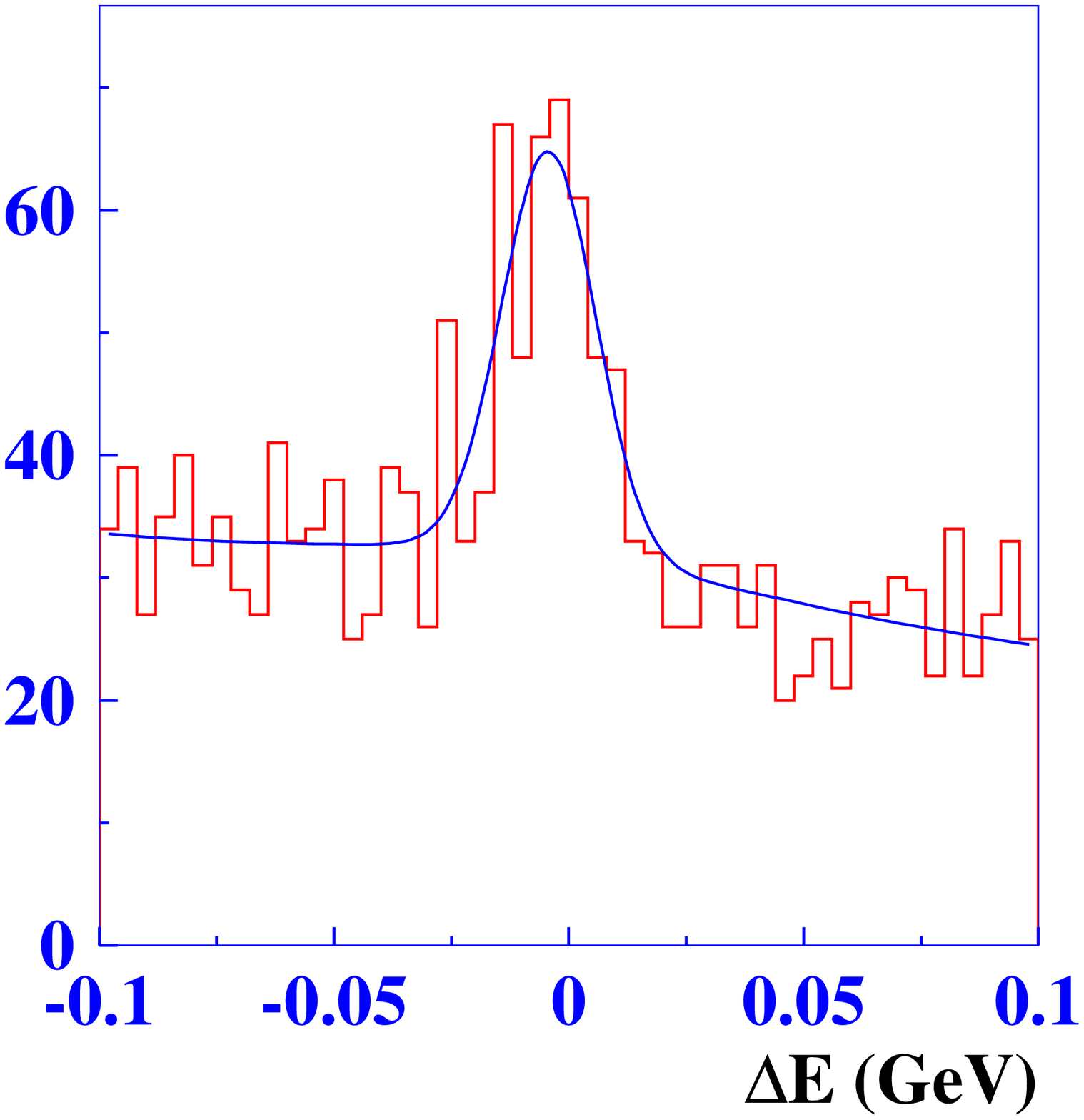}\\
\includegraphics[height=8 cm]{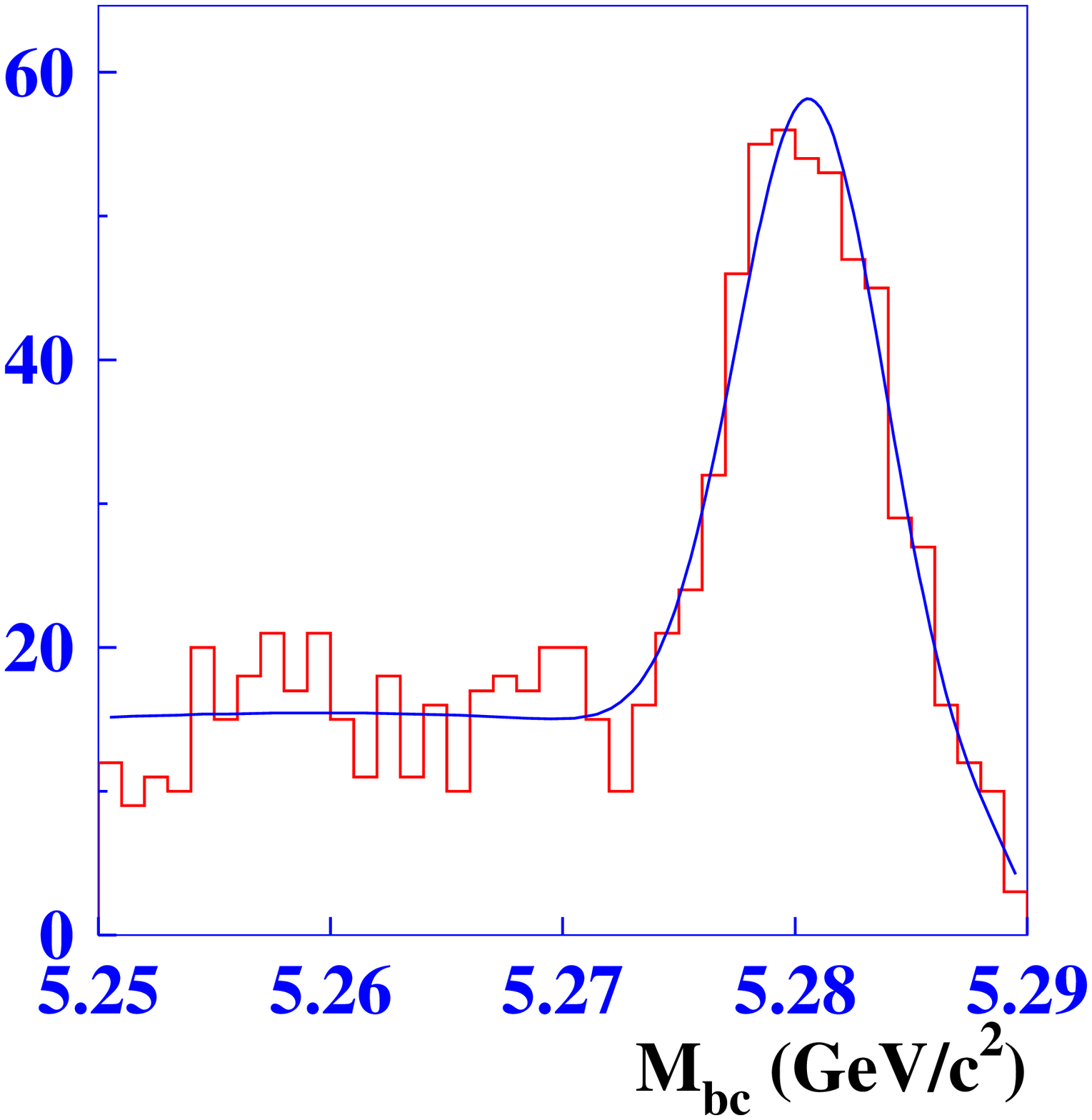}&
\includegraphics[height=8 cm]{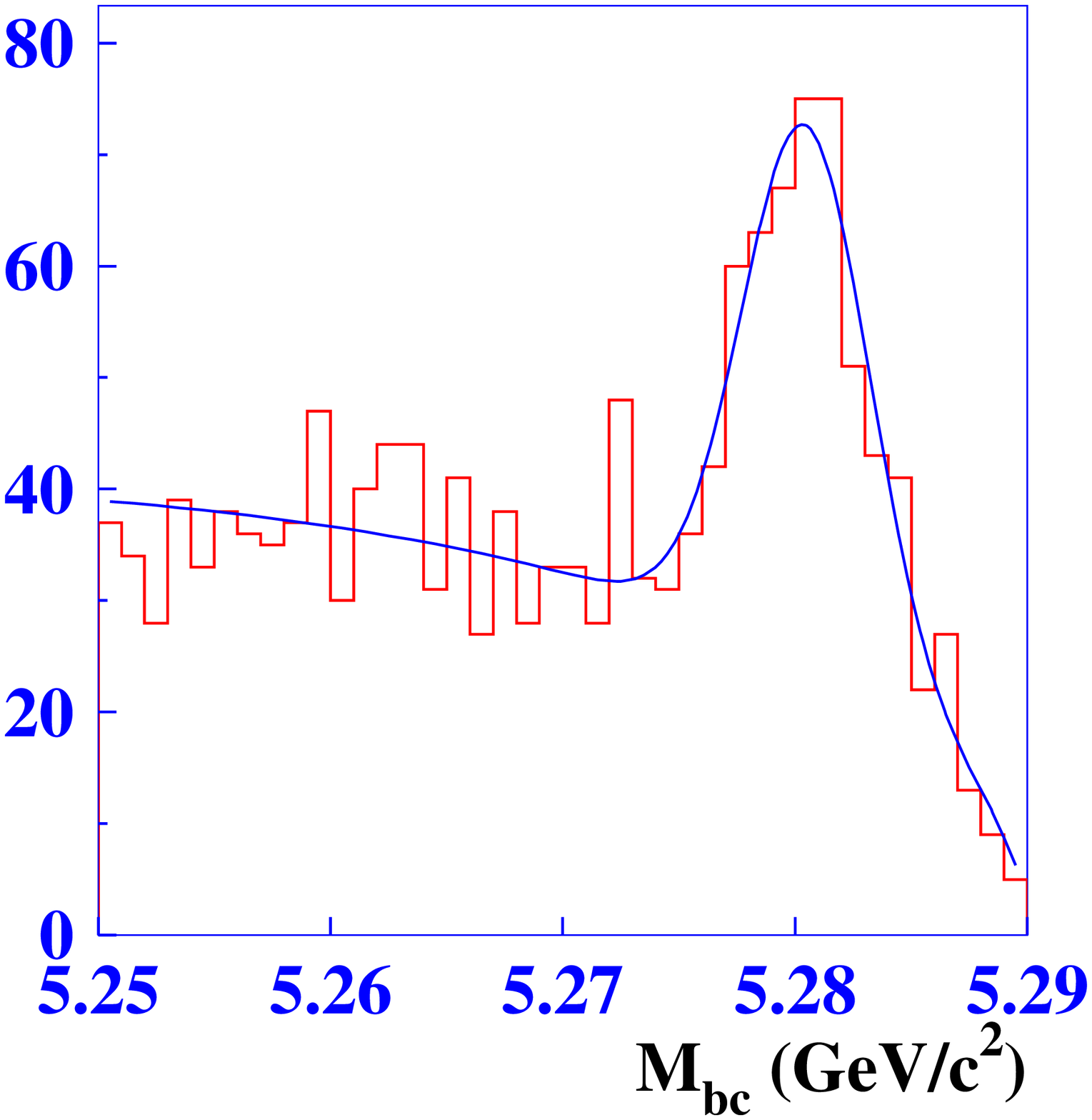}\\
\vspace*{-16 cm} & \\
{\bf\large \hspace*{-3cm} a)}&{\bf\large \hspace*{-3cm} b)}\\
\vspace*{7. cm} & \\
{\bf\large \hspace*{-3cm} c)}&{\bf\large \hspace*{-3cm} d)}\\
\vspace*{6.5 cm} & \\
$D\to K\pi$&$D\to K3\pi$\\
\end{tabular}
\caption{Distribution of $\bar{B}^{0}\to D^{*0}\pi^{+}\pi^{-}$ events
in $\Delta E$ and $M_{\rm bc}$ plots with $D$ reconstruction in $K\pi$
mode -- (a), (c) and in $K3\pi$ mode -- (b), (d). 
(a), (b) $\Delta E$  distributions.
(c), (d) $M_{\rm bc}$ distributions.}
\label{f:1mb}
\end{center}
\end{figure}

For $D^*$ reconstruction,  the $D^{*0}\to D^0\pi^0$ decay is used with two  
decay modes of the $D^0$: $D^0\to K^-\pi^+$ and 
$D^0\to K^-\pi^+\pi^+\pi^-$.
The event distributions in $\Delta E$ and $M_{bc}$  are shown 
in Fig.~\ref{f:1mb}.
In each mode the number of signal events is obtained in a way  similar to 
that described for the $D\pi\pi$ selection. 
The observed signal yields of
$N_{K\pi}=278\pm23$ and $N_{K3\pi}=269\pm29$ for the $K\pi$ and $K\pi\pi\pi$
modes,
respectively, are consistent when efficiencies determined 
from MC, and the following $D$
branching fractions are used: 
($3.80\pm 0.09)\%$ for $K^-\pi^+$ and 
$(7.46\pm 0.31)\%$ for $K^-\pi^+\pi^+\pi^-$.
The branching fraction of $(D^*\to D\pi)\pi\pi$ events, 
calculated from the 
weighted average of the values obtained for the two modes, is:
$$
{\cal B}(\bar{B}^0\to D^{*0}\pi^+\pi^-)=(1.09\pm0.08\pm0.16)\times10^{-3},
$$
where the first error is statistical and the second is systematic.
This  measurement is about $2.5\,\sigma$ larger than the  previous 
result~\cite{Asish} of
$
(0.62\pm0.22\pm0.22)\times 10^{-3}. 
$
We consider the previous measurement to be a statistical fluctuation.
The systematic error contributions are listed in Table~\ref{t:syss}.

The background shape uncertainty and the effect of cut boundaries were
estimated in the same way
as for the $D\pi\pi$ analysis.
\begin{table}
\begin{center}
\begin{tabular}{l|l}
\hline
Source &{$\sigma_{sys},~\%$}\\
\hline
$Br({D^{*0},~D^0})$& 5.3\\
Tracking&4.3\\
$\pi^0$&6\\
PID &5\\
MC&3\\
Background&10\\
\hline
Total&14.7\\
\hline
\end{tabular}
\end{center}
\caption{Contributions to the systematic error for $B\to D^*\pi\pi$.} 
\label{t:syss}
\end{table}

\subsection{$B\to D^*\pi\pi$  coherent amplitude analysis}
In this final state we have a decaying vector $D^*$ particle.  
There are two additional degrees of freedom and, 
in addition to the $D^*\pi$ and $\pi\pi$ 
invariant masses squared ($q^2,q^2_1$),  two other variables
are needed to specify the final state.
The variables are chosen to be
the angle $\alpha$ between
the pions from the  $D^{**}$ and  $D^*$ decay in the 
$D^*$ rest frame, and the azimuthal angle $\gamma$ 
of the pion from the  $D^{*}$ 
relative to the $B\to D^*\pi\pi$ decay plane.
For the case of the $\pi\pi$ structure analysis another set of
parameters can be chosen: $\pi\pi$ square mass ($q_1^2$); the helicity
angle $\theta'$ of
the $\pi\pi$-meson --- the angle between  the positive pion from this
meson decay 
 and  the $D^*$  direction in the $\pi\pi$ meson rest frame; the
 helicity angle $\alpha'$  of the $D^*$ meson --- the angle between  the pion from the $D^*$ 
decay 
 and the $\pi\pi$-meson    in the  $D^*$  rest frame; and the angle $\gamma'$
 between the decay planes of the $D^*$ and the $\pi\pi$-meson.

For further analysis, events satisfying  
the selection criteria described in the first section
within the signal region
$((\Delta E+\kappa(M_{\rm bc}-M_B))/\sigma_{\Delta E})^2+((M_{\rm
  bc}-M_B)/\sigma_{M_{\rm bc}})^2<s$ 
are selected. The parameters $\sigma_{\Delta E}=11
{\rm MeV}/c^2,~\sigma_{M_{\rm bc}}=2.7~{\rm MeV}/c^2,~\kappa=0.9$  are
obtained from a fit to experimental data, and the coefficient
$\kappa$ takes into account the  correlation between $M_{\rm bc}$ and $\Delta E$.
The parameter $s$ is selected to be 4 for the mode with $D\to K\pi$
and 3 for $D\to K\pi\pi\pi$ to have a similar signal-to-background ratio.
To understand the contribution and shape of the background, we use
events in the
sidebands 
$((\Delta E\pm65\,{\rm MeV}+\kappa(M_{\rm bc}-M_B))/\sigma_{\Delta E})^2+((M_{\rm
  bc}-M_B)/\sigma_{M_{\rm bc}})^2<s$.

\begin{figure}[h]
\begin{center}
\begin{tabular}{cc}
\includegraphics[height=8 cm]{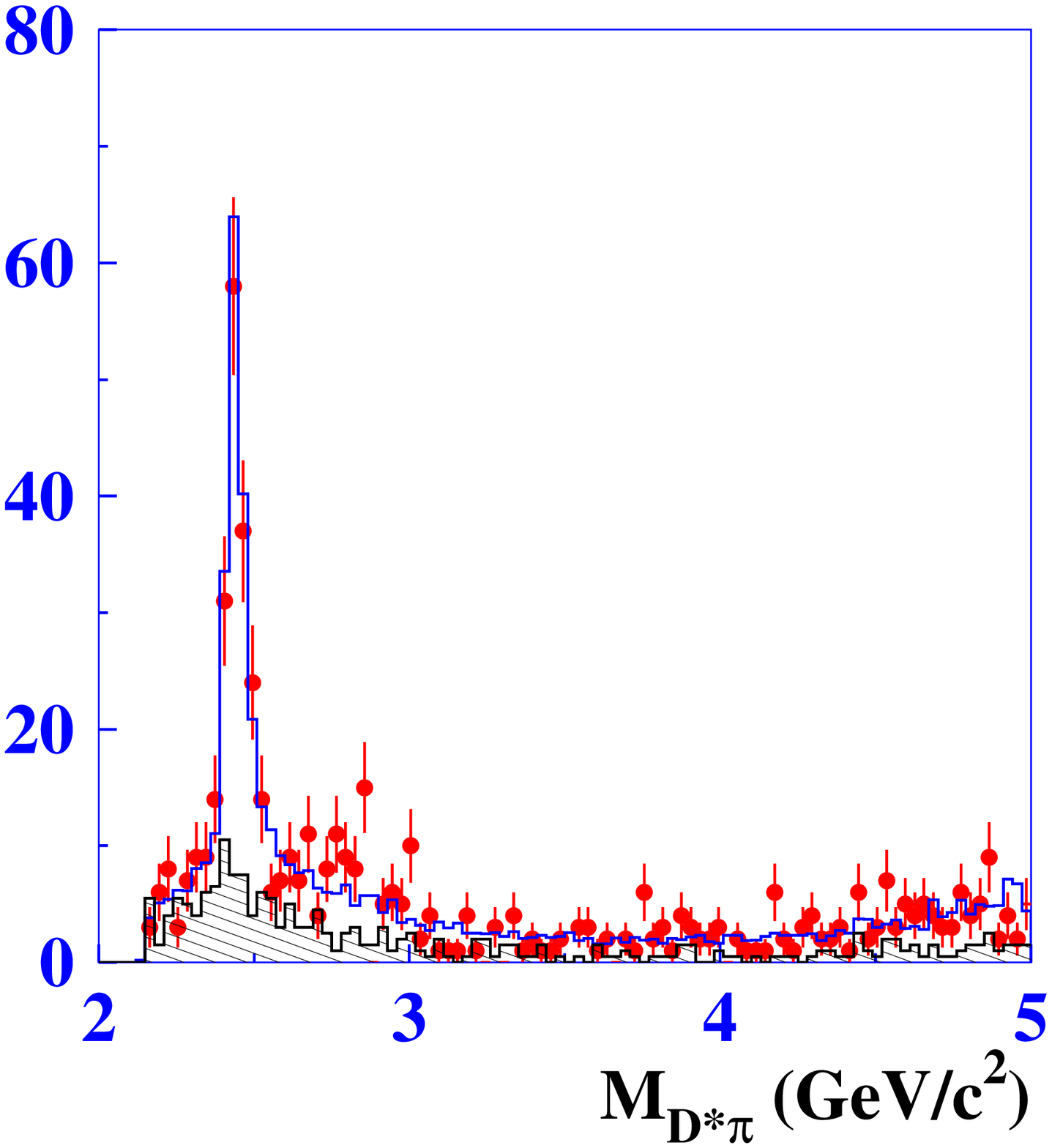}&
\includegraphics[height=8 cm]{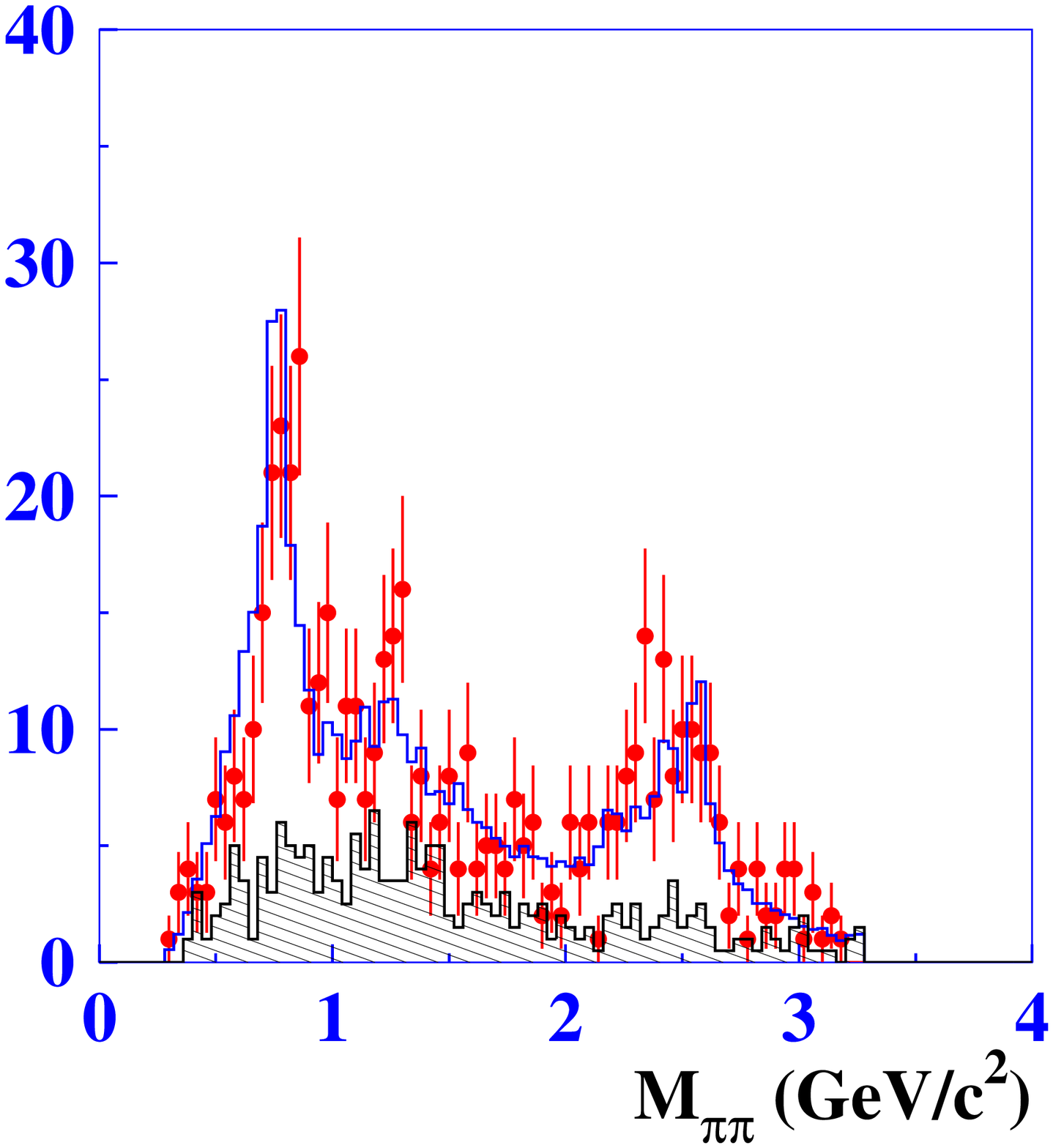}\\
\vspace*{-8 cm} & \\
{\bf\large \hspace*{-3cm} a)}&{\bf\large \hspace*{-3cm} b)}\\
\vspace*{6.5 cm} & \\
\end{tabular}
\caption{Mass distribution of $D^*\pi$ and $\pi\pi$ events. Points are the experimental data, the hatched histogram
is the background distribution obtained from the sidebands, the open histogram is MC simulation 
with the amplitudes and parameters of the intermediate resonances obtained from the fit.}
\label{f:dsp_M}
\end{center}
\end{figure}

\begin{figure}[h]
\begin{center}
\begin{tabular}{cc}
\includegraphics[height=8 cm]{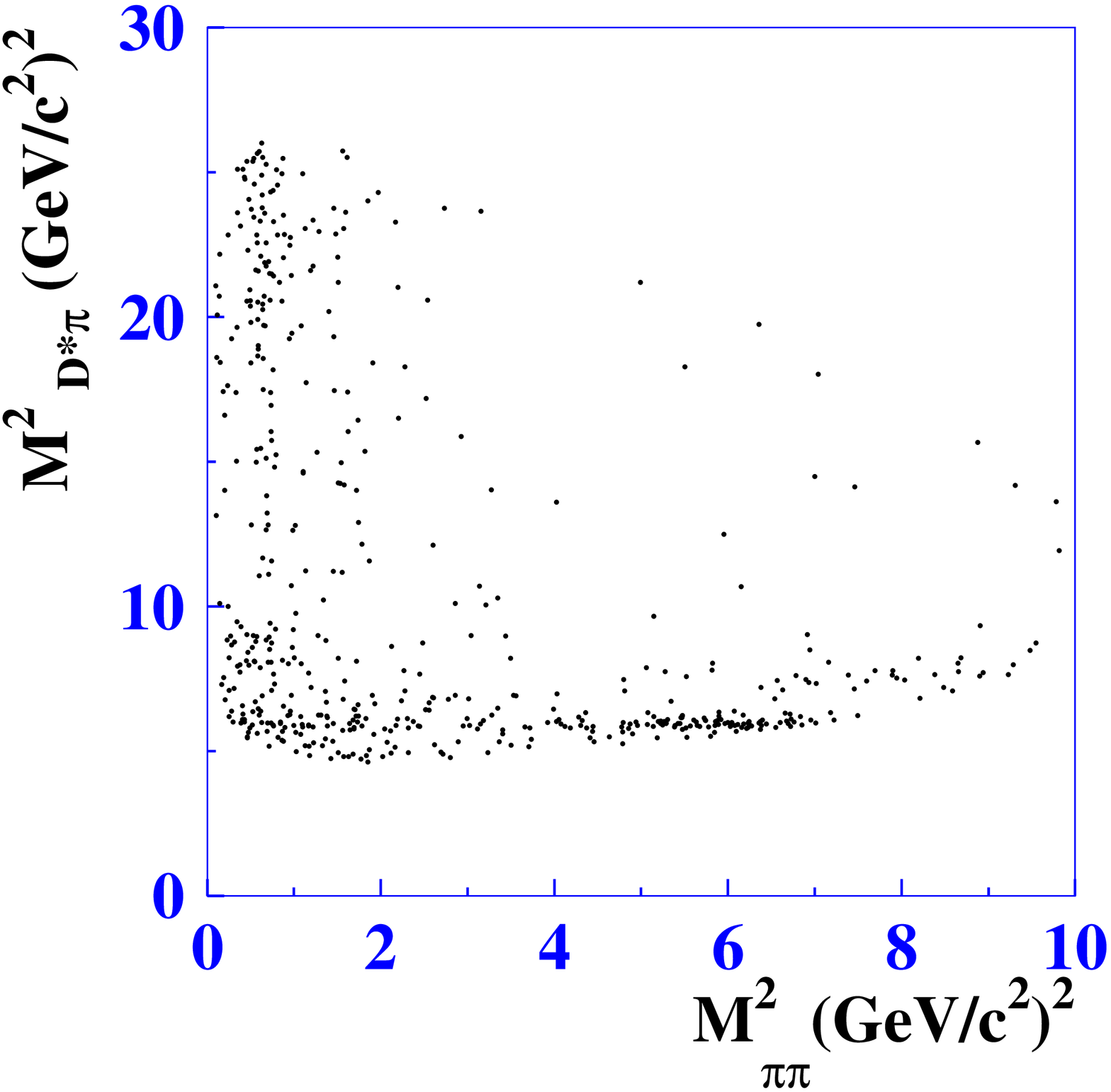}&
\includegraphics[height=8 cm]{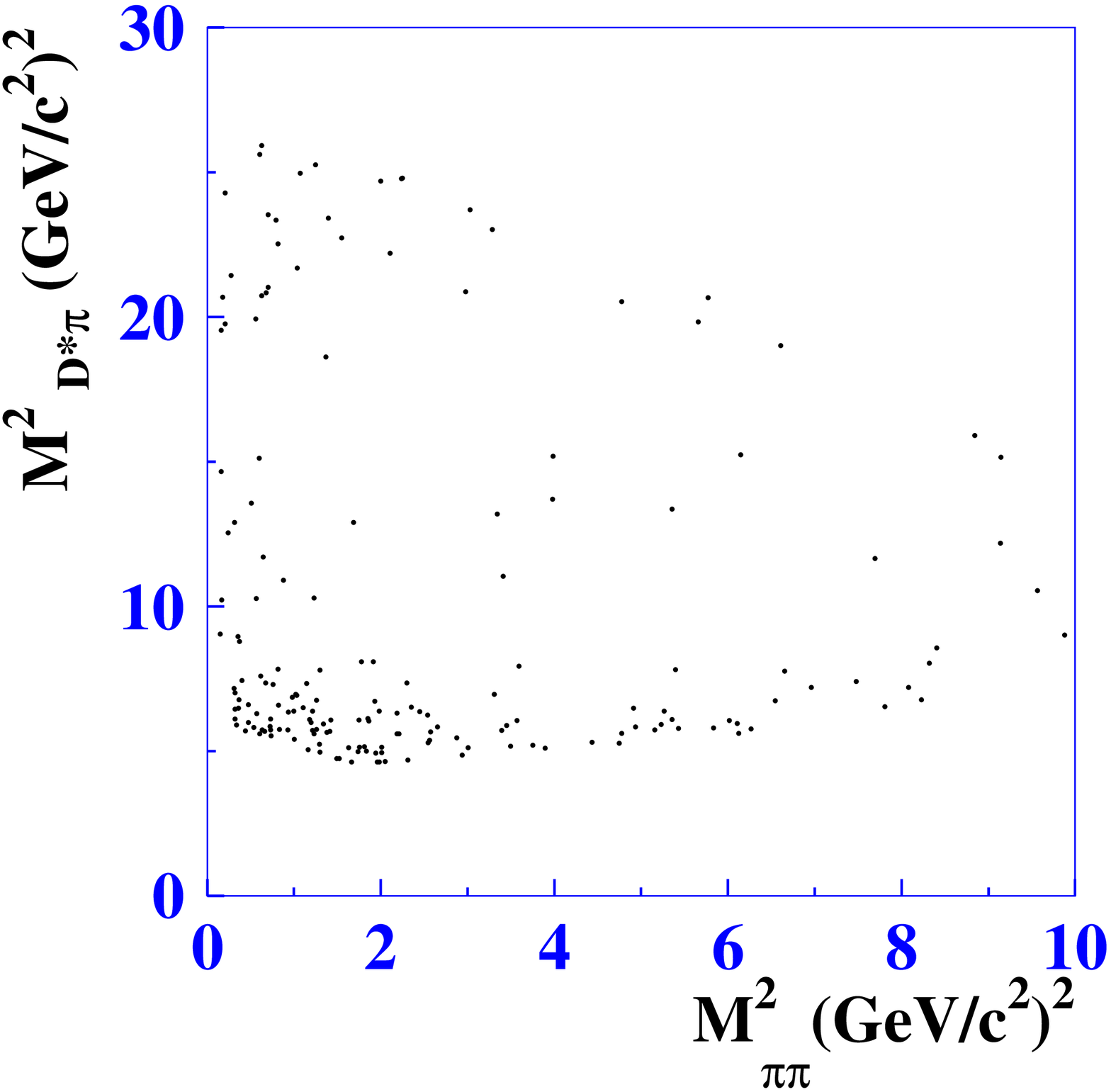}\\
\vspace*{-8 cm} & \\
{\bf\large \hspace*{-3cm} a)}&{\bf\large \hspace*{-3cm} b)}\\
\vspace*{6.5 cm} & \\
\end{tabular}
\caption{The Dalitz plot of (a) signal events; (b) sideband events.}
\label{f:dpp_DS}
\end{center}
\end{figure}

The $D^*\pi$ and $\pi\pi$ mass distributions for the signal and sideband events 
are shown in Fig.~\ref{f:dsp_M} and \ref{f:dpp_DS}. There is a  clear peak of the narrow states
$D_2^{*+}$ and $D_1^{+}$ with a negligible contribution of the broad
states in the $D\pi$ distribution. In the $\pi\pi$ distribution the peaks of
$\rho$ and $f_2(1270)$ are clearly seen while the peak-like structure around
2.6~GeV/c$^2$  is the reflection of the $D\pi$ angular distribution.
In order to have the same Dalitz plot boundary for 
events  from both the signal and the sideband regions as well as to decrease the 
smearing effect introduced by the detector resolution,
mass-constrained fits of $D\pi$ to $M_{D^{*+}}$ 
and $D^*\pi\pi$ to $M_B$ are performed. 

To extract the amplitudes and phases for different intermediate states, 
an  unbinned  likelihood  fit in 
the four-dimensional phase space  is performed. 
Assuming that the background distribution ${\cal
  B}(q^2,q^2_1,\alpha,\gamma)$
in the signal region  has the 
same shape as in the $\Delta E$ sideband,
we obtain the 
${\cal B}(q^2,q^2_1,\alpha,\gamma)$ dependence from a fit 
of the sideband 
distribution to a smooth four-dimensional function.

The number of background events in the signal region 
is normalized according to the relative areas of the signal and the 
sideband regions. 
The signal is parameterized as a sum of the 
amplitudes of an intermediate tensor ($D_2^*$), two axial vector 
mesons ($D'_1,~D_1$), and three resonances $\rho,~f_2$ and $f_0(600)$ 
in the $\pi\pi$ mode. For $\rho$ and $f_2$ there can be three
different amplitudes depending on the relative polarizations of the
decay products:$A_{0},~A_{\perp}$ and  $A_{||}$. 

Finally, the signal is expressed as follows:
\begin{eqnarray}
\label{e:4.1}
&&S(q^2,q_1^2,\alpha,\gamma)=|a_2A^{(2)}(q^2,q_1^2,\alpha,\gamma)+
a_1e^{i\phi_1}A^{(n)}(q^2,q_1^2,\alpha,\gamma)
+a_we^{i\phi_w}A^{(w)}(q^2,q_1^2,\alpha,\gamma)\nonumber\\
&&+
a_{\rho}e^{i\phi_{\rho}}((1-a_{||}^{\rho}-a_{\perp}^{\rho})A^{(\rho)}_{0}(q^2,q_1^2,\alpha,\gamma)+a_{||}^{\rho}e^{i\phi^{\rho}_{||}}A^{(\rho)}_{||}(q^2,q_1^2,\alpha,\gamma)\nonumber\\
&&+a_{\perp}^{\rho}e^{i\phi^{\rho}_{\perp}}A^{(\rho)}_{\perp}(q^2,q_1^2,\alpha,\gamma))\nonumber\\
&&+
a_{f_2}e^{i\phi_{f_2}}((1-a_{||}^{f_2}-a_{\perp}^{f_2})A^{(f_2)}_{0}(q^2,q_1^2,\alpha,\gamma)+a_{||}^{f_2}e^{i\phi^{f_2}_{||}}A^{(f_2)}_{||}(q^2,q_1^2,\alpha,\gamma)\nonumber\\
&&+a_{\perp}^{f_2}e^{i\phi^{f_2}_{\perp}}A^{(f_2)}_{\perp}(q^2,q_1^2,\alpha,\gamma))\nonumber\\
&&+
a_{f_0}A^{(f_0)}(q^2,q_1^2,\alpha,\gamma)+
a_{ps}|^2.
\end{eqnarray}
The MC gives the resolution in invariant mass of about $1.9$~MeV/$c^2$,  
which is smaller than the resonance widths and can be neglected.

Each resonance is described by a relativistic Breit-Wigner function 
with a width 
depending on $q^2$.
The angular dependence for each resonance corresponds
to the spins of the intermediate and final state
particles~\cite{mybelle}.
The amplitudes of $B\to D^*\rho$ and $B\to D^*f_2$ decays have
the following angular distributions:
\begin{eqnarray}
\label{e:3.4}
A^{\rho}_{0}&\sim&\cos{\alpha'}\cos{\theta'}\nonumber\\
A^{\rho}_{||}&\sim&\sin{\alpha'}\sin{\theta'}\cos{\gamma'}\nonumber\\
A^{\rho}_{\perp}&\sim&\sin{\alpha'}\sin{\theta'}\sin{\gamma'}
\end{eqnarray}
\begin{eqnarray}
\label{e:3.4}
A^{f_2}_{0}&\sim&\cos{\alpha'}(\cos^2{\theta'}-1/3)\nonumber\\
A^{f_2}_{||}&\sim&\sin{\alpha'}\sin{\theta'}\cos{\theta'}\cos{\gamma'}\nonumber\\
A^{f_2}_{\perp}&\sim&\sin{\alpha'}\sin{\theta'}\cos{\theta'}\sin{\gamma'}
\end{eqnarray}

Table~\ref{t:dpps1} shows the results of the fit for different models. 
If we remove the $D'_1$ meson from Eq.~\ref{e:4.1}, the likelihood
   does not change significantly.  If we remove the $f_2$ instead, the
   likelihood increases by 41.
Adding the phase space
term does not improve the likelihood significantly.  Replacing the  $f_2$ 
meson with an $f_0$(1370) results in a worse likelihood value as shown 
in Table~\ref{t:dpp1s}.
\begin{table}
\begin{tabular}{|c|c|c|c|}
\hline
&\multicolumn{3}{|c|}{$D^*_2,~D_1,~D'_1,~\rho,~f_0(600)$}\\
\hline
&$f_2$&$f_0$(1370)& no\\
\hline
$ -2\ln{\cal L}/{\cal L}_0$& 0 &11&41\\
\hline
\end{tabular}
\caption{Comparison of the models with different resonances included in the fits.}
\label{t:dpp1s}
\end{table}

The masses and widths of the $\pi\pi$ resonances are fixed at their PDG
values; $M_{D_2^{*+}}$, and $\Gamma^0_{D_2^{*+}}$ are taken from
the $D\pi\pi $ fit; and  $ M_{D'_1}=2427\,{\rm MeV}/c^2,~\Gamma_{D'_1}^0=384\,{\rm MeV}/c^2$ have been taken from our
measurement for $D^{**0}$~\cite{mybelle}.
The mass $M_{D_1^+}$ and width $\Gamma_{D_1^+}$ as well as the branching fractions and phases of the amplitudes  were free parameters of the 
fit. 

\begin{table}
 \begin{tabular}{|c|c|c|c|c|}
 \hline
 \hline
 & $D^*_2,~D_1,~D'_1,$ &
 $D^*_2,~D_1,$
 &$D^*_2,~D_1,~D'_1,$       & $D^*_2,~D_1,~D'_1,$       
 \\
&&&&\\
 & $\rho,~f_2,~f_0(600)$ &
 $\rho,~f_2,~f_0(600)$
 &$\rho,~f_0(600)$       & $\rho,~f_2,~f_0(600)+ps$       \\
 \hline
$ -2\ln{\cal L}/{\cal L}_0$ &  0    &  -4    &  +41    &  -1    \\
 \hline
$Br_{D_2^*}(10^{-4})$   &   2.45$\pm$ 0.42&   2.45$\pm$ 0.42&   2.48$\pm$ 0.43&   2.43$\pm$ 0.41\\
 \hline
$\phi_{D_1}$   &  0.908$\pm$0.145&  0.907$\pm$0.145&  0.837$\pm$0.139&  0.766$\pm$0.147\\
$Br_{D_1}(10^{-4})$     &   3.68$\pm$ 0.60&   3.71$\pm$ 0.62&   4.03$\pm$ 0.84&   3.63$\pm$ 0.61\\
 \hline
$\phi_{D'_1}$  & -0.197$\pm$0.584&  --& -0.316$\pm$0.670& -0.121$\pm$0.556\\
$Br_{D'_1}(10^{-4})$    &   0.14$\pm$ 0.13&   --&   0.11$\pm$ 0.12&   0.14$\pm$ 0.14\\
 \hline
$\phi_{ps}$ &  --&  --&  --&  2.594$\pm$0.551\\
$Br_{ps}(10^{-4})$   &   --&   --&   --&   0.00$\pm$ 0.17\\
 \hline\hline
$\phi_{\rho}$  &  2.566$\pm$0.333&  2.543$\pm$0.337&  2.032$\pm$0.326&  2.560$\pm$0.327\\
$Br_{\rho}(10^{-4})$    &   3.73$\pm$ 0.87&   3.78$\pm$ 0.87&   3.89$\pm$ 0.96&   3.74$\pm$ 0.85\\
 \hline
$\phi_{f_2}$   &  0.440$\pm$0.413&  0.411$\pm$0.425&  --&  0.429$\pm$0.453\\
$Br_{f_2}(10^{-4})$     &   1.05$\pm$ 0.37&   1.05$\pm$ 0.37&   --&   0.98$\pm$ 0.35\\
 \hline
$\phi_{f_0}$   & -2.263$\pm$0.646& -2.190$\pm$0.643& -2.823$\pm$0.498& -2.181$\pm$0.597\\
$Br_{f_0}(10^{-4})$     &   0.17$\pm$ 0.11&   0.16$\pm$ 0.11&   0.32$\pm$ 0.17&   0.17$\pm$ 0.11\\
 \hline\hline
$a^{\rho}_{||}$      &  0.204$\pm$0.059&  0.198$\pm$0.058&  0.176$\pm$0.061&  0.211$\pm$0.059\\
$a^{\rho}_{\perp}$   &  0.067$\pm$0.038&  0.065$\pm$0.038&  0.105$\pm$0.042&  0.066$\pm$0.038\\
$\phi^{\rho}_{\perp}$&  0.678$\pm$0.348&  0.686$\pm$0.351&  0.693$\pm$0.307&  0.624$\pm$0.358\\
$a^{\rho}_{0}$    &  0.730$\pm$0.058&  0.737$\pm$0.057&  0.719$\pm$0.059&  0.723$\pm$0.058\\
$\phi_{\rho_2}$      &  2.046$\pm$0.229&  2.031$\pm$0.229&  2.269$\pm$0.250&  2.030$\pm$0.225\\
 \hline\hline
$a^{f_2}_{0}$     &  0.623$\pm$0.137&  0.616$\pm$0.143&  --&  0.646$\pm$0.142\\
$a^{f_2}_{\perp}$    &  0.080$\pm$0.084&  0.092$\pm$0.091&  --&  0.082$\pm$0.085\\
$\phi^{f_2}_{\perp}$ & -3.036$\pm$0.687& -2.983$\pm$0.672& --& -3.129$\pm$0.752\\
$a^{f_2}_{||}$       &  0.297$\pm$0.137&  0.292$\pm$0.142&  --&  0.273$\pm$0.143\\
$\phi^{f_2}_{||}$    & -0.895$\pm$0.489& -0.846$\pm$0.506& --& -0.926$\pm$0.551\\
 \hline\hline
 \end{tabular}
\caption{The fit results for different sets of amplitudes.}
\label{t:dpps1}
\end{table}
In addition to the main minimum, there are several local minima that 
differ in $-2\ln{\cal L}/{\cal L}_0$ by 2.1 to 25. The main minimum and local
minima are treated in the same way as for $D\pi\pi$.

\begin{figure}[h]
\begin{tabular}{ccc}
\includegraphics[height=5 cm]{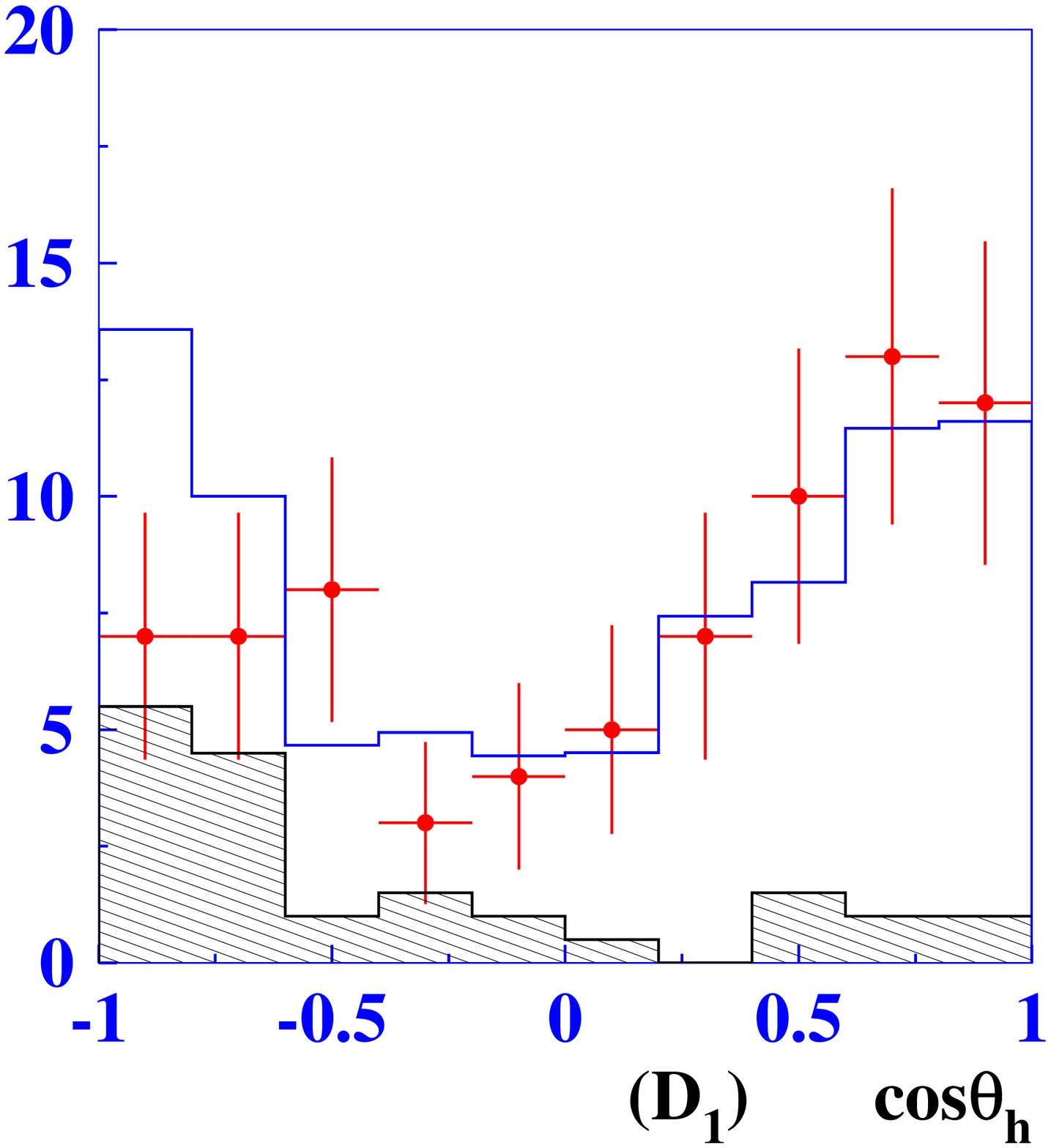}&
\includegraphics[height=5 cm]{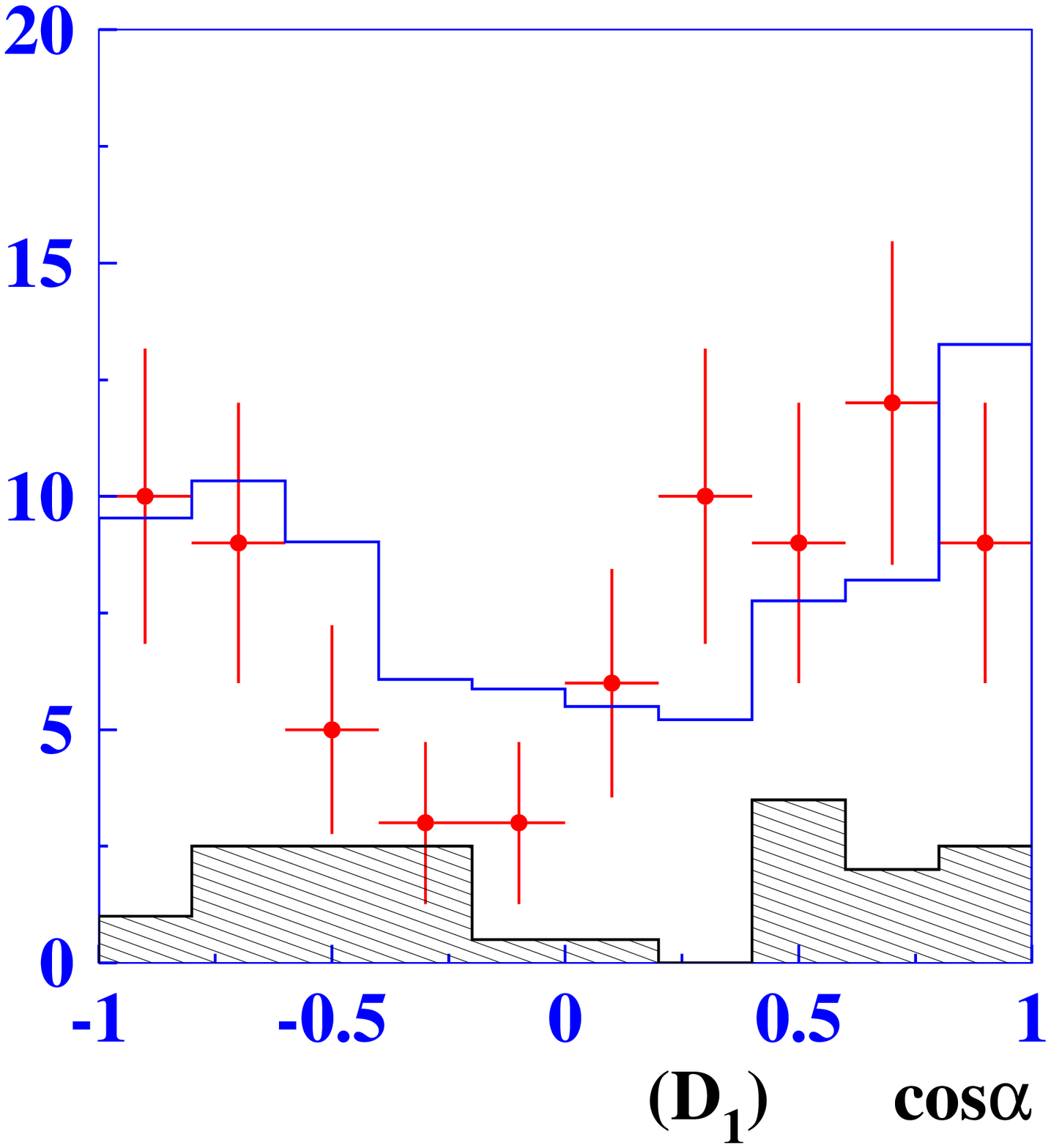}&
\includegraphics[height=5 cm]{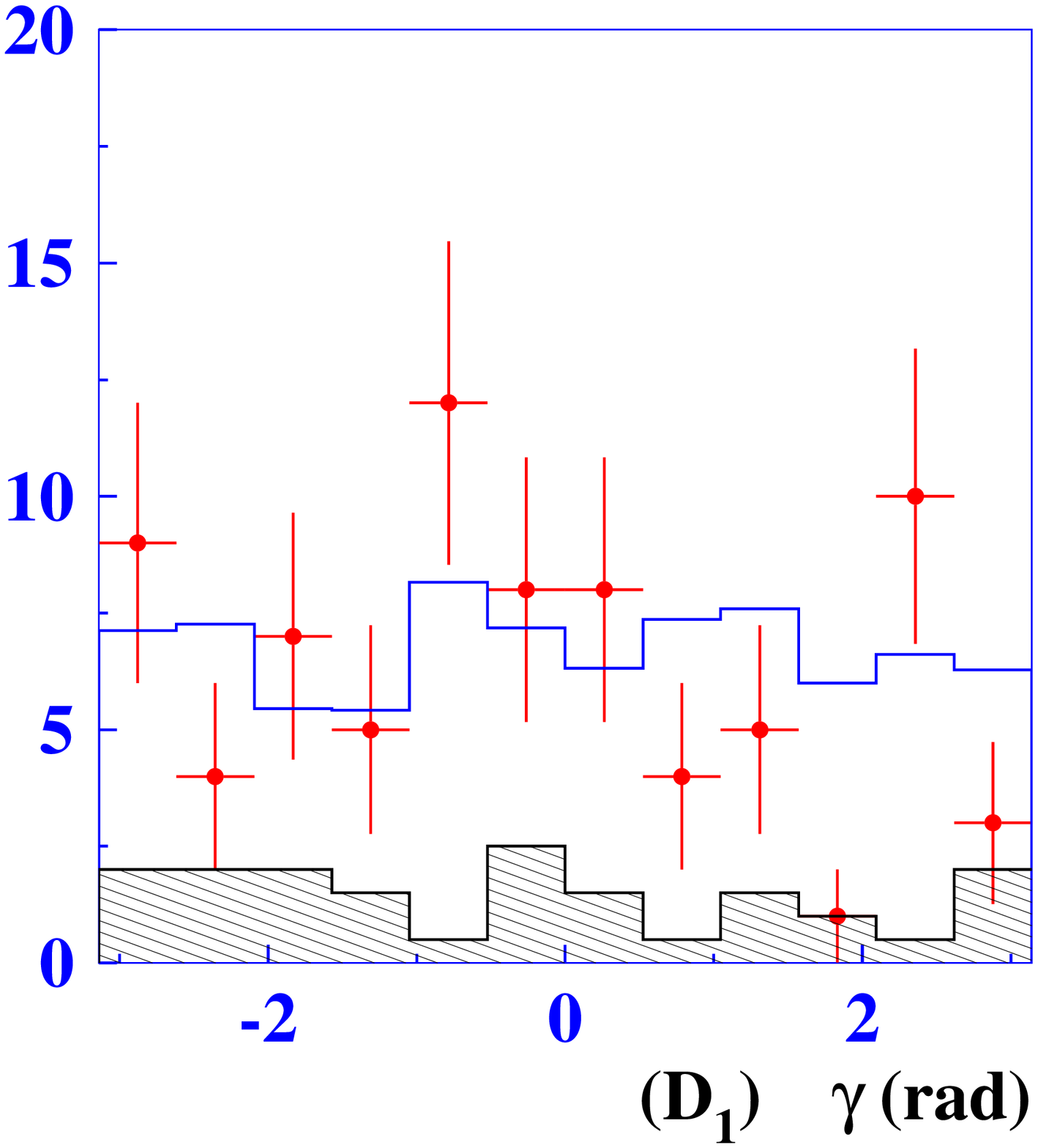}\\
\includegraphics[height=5 cm]{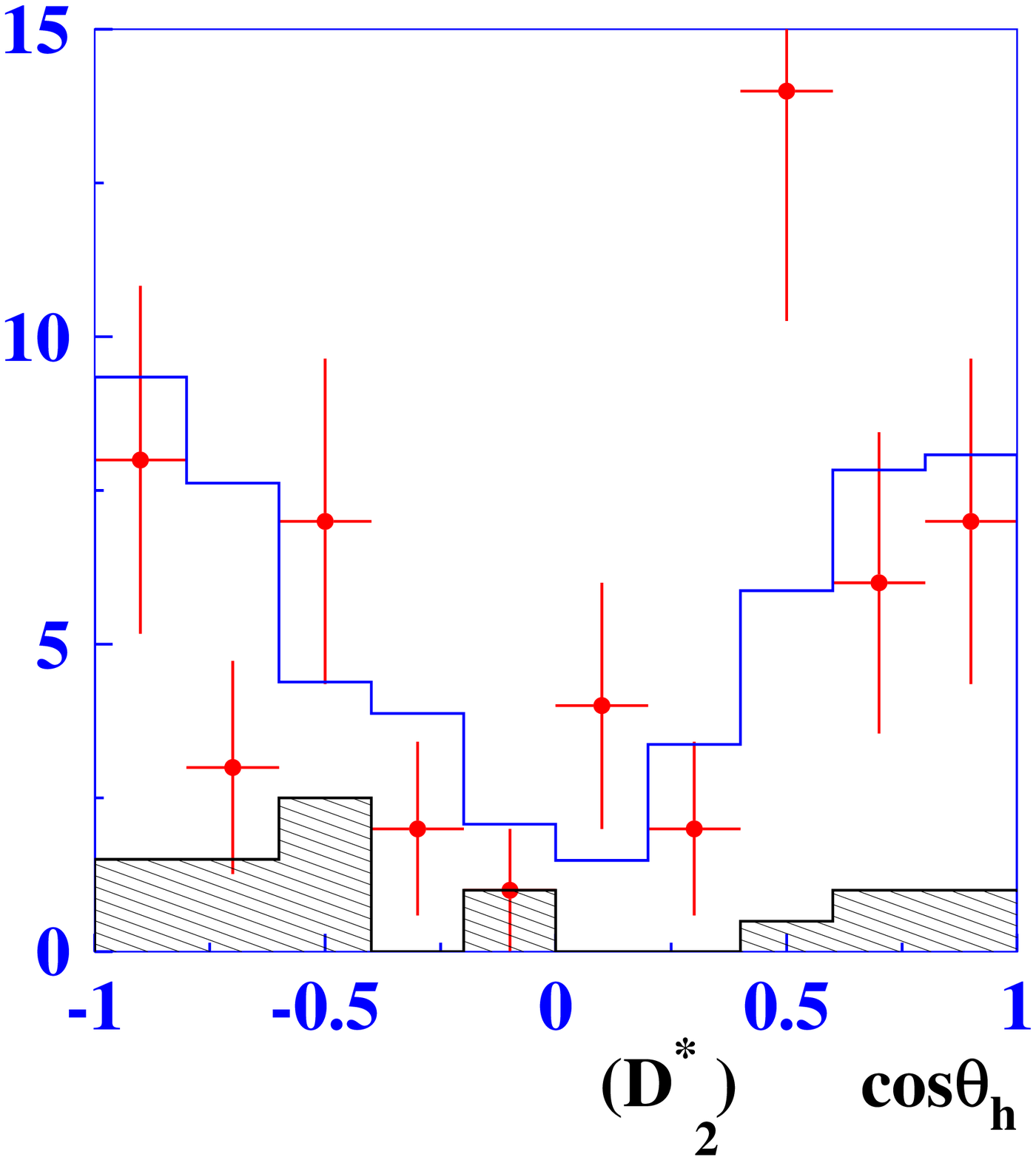}&
\includegraphics[height=5 cm]{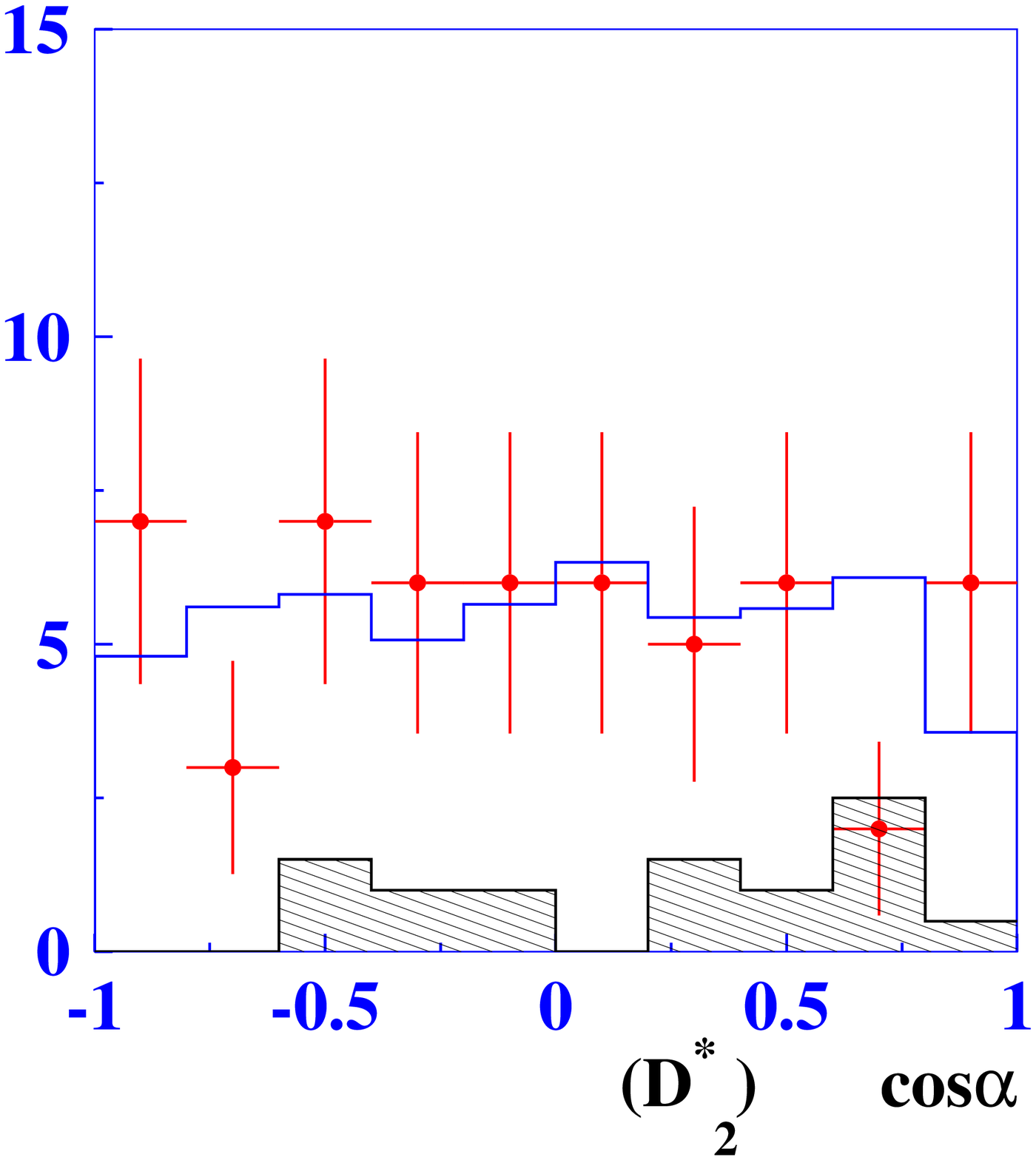}&
\includegraphics[height=5 cm]{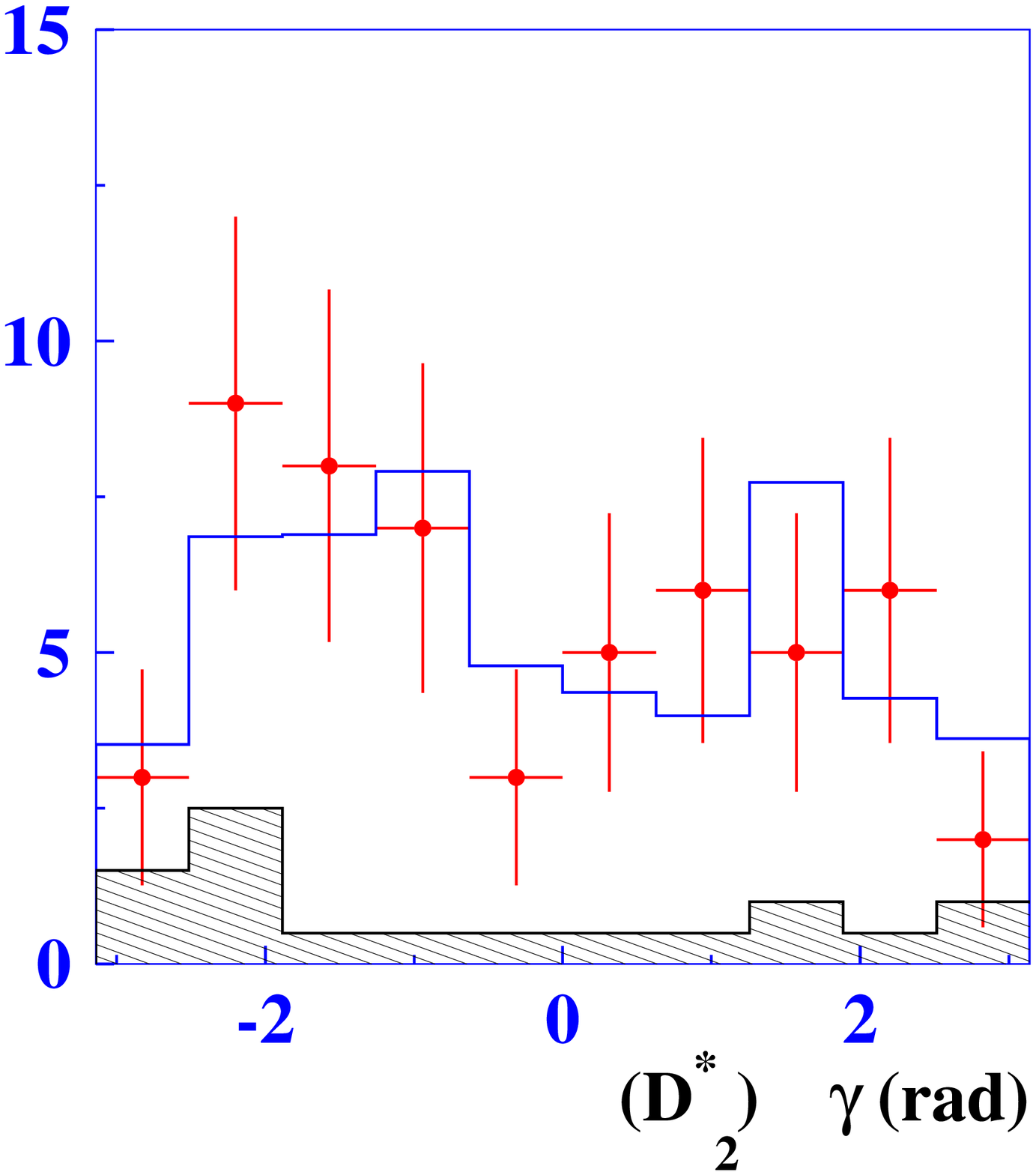}\\
\vspace*{-10 cm} & \\
{\bf\large \hspace*{-1cm} a)}&{\bf\large \hspace*{-1cm} b)}&{\bf\large \hspace*{-1cm} c)}\\
\vspace*{4 cm} & \\
{\bf\large \hspace*{-1cm} d)}&{\bf\large \hspace*{-1cm} e)}&{\bf\large \hspace*{-1cm} f)}\\
\vspace*{3 cm} & \\
\end{tabular}
\caption{The distribution of the data for the $D_1$ region: (a)-(c)
  $|M_{D^*\pi}-2.41|<0.03\,GeV/c^2$ and the $D_2^*$ region:
  $|M_{D^*\pi}-2.45|<0.025\,GeV/c^2$ (d)-(f).
(a), (d) $\cos\theta$ -- helicity angle of $D^{**}$; 
(b), (e) $\cos\alpha$ -- helicity angle of $D^{*}$; 
(c), (f) azimuthal angle $\gamma$.
The points are experimental data, the histogram is MC with fitted parameters,
and the hatched histogram is the background contribution (from sideband).}
\label{f:D1}
\end{figure}

Figs.~\ref{f:dsp_M} and~\ref{f:D1}   demonstrate  
the comparison of data and the MC simulation generated  
according to Eq.~(\ref{e:4.1}) with parameters obtained from the fit.
The distributions of the helicities of 
 $D^{**}$ and $D^{*}$ and the angle $\gamma$ in different  $q^2$ regions 
demonstrate reasonable agreement  of the experimental data and fit result.

For  the $D_1$ meson we obtain the following parameters:
$$
M_{D^{+}_1}=(2428.2\pm2.9\pm1.6\pm0.6)\, {\rm MeV}/c^2,~\Gamma_{D^{+}_1}=(34.9\pm6.6^{+4.1}_{-0.9}\pm4.1)\, \rm{MeV}.
$$
These parameters  are in good agreement with the CLEO measurement for $D^0_1$:\\
$
M_{D^{0}_1}=(2425\pm2\pm2)\, {\rm MeV}/c^2,~\Gamma_{D^{0}_1}=(26^{+8}_{-7}\pm4)\, \rm{MeV}
$~\cite{dobs}.

The preliminary results for the product of the branching ratios of the  
$D^{**}$'s are the following:
$$
{\cal B}(\bar{B}^0\to D_1^+\pi^-)\times B(D_1^+\to D^{*0}\pi^+)=(3.68\pm0.60^{+0.71+0.65}_{-0.40-0.30})\times10^{-4},
$$
$$
{\cal B}(\bar{B}^0\to D^{*+}_2\pi^-)\times B(D_2^{*+}\to D^{*0}\pi^+)=(2.45\pm0.42^{+0.35+0.39}_{-0.45-0.17})\times10^{-4},
$$
$$
{\cal B}(\bar{B}^0\to D'^{+}_1\pi^-)\times B(D'^{+}_1\to D^{*0}\pi^+)=(0.14\pm0.13\pm0.12^{+0.00}_{-0.10})\times10^{-4}.
$$
The last value is not statisticaly significant and  corresponds to an upper limit: 
$$
{\cal B}(\bar{B}^0\to D'^{+}_1\pi^-)\times B(D'^{+}_1\to
D^{*0}\pi^+)<0.7\times10^{-4} \rm~at~90\,\%~C.L.
$$
Including a contact term improves the likelihood but without high significance (see Table~\ref{t:dpps1}).

\begin{figure}[h]
\begin{center}
\begin{tabular}[c]{ccc}
\includegraphics[height=5 cm]{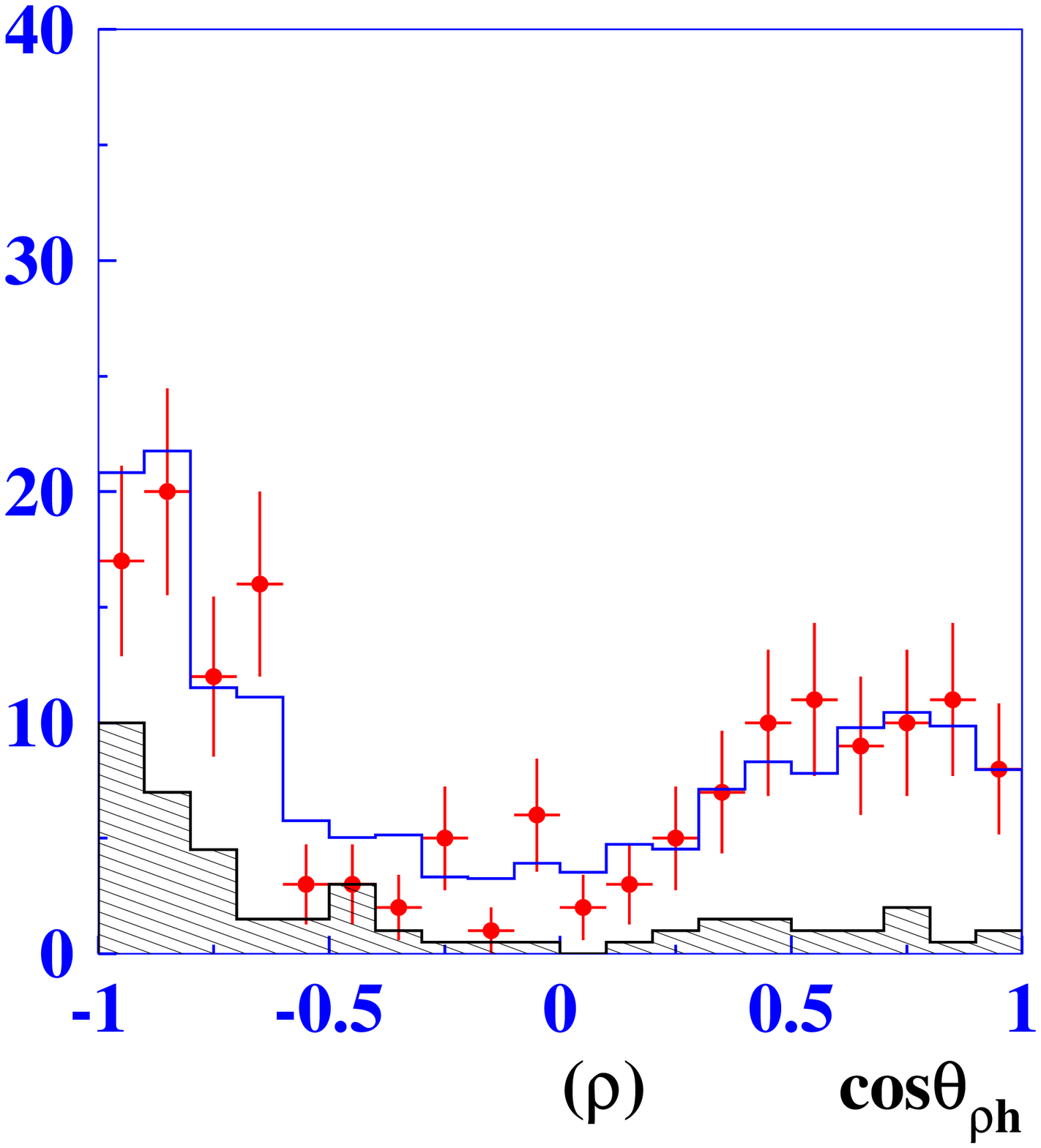}&
\includegraphics[height=5 cm]{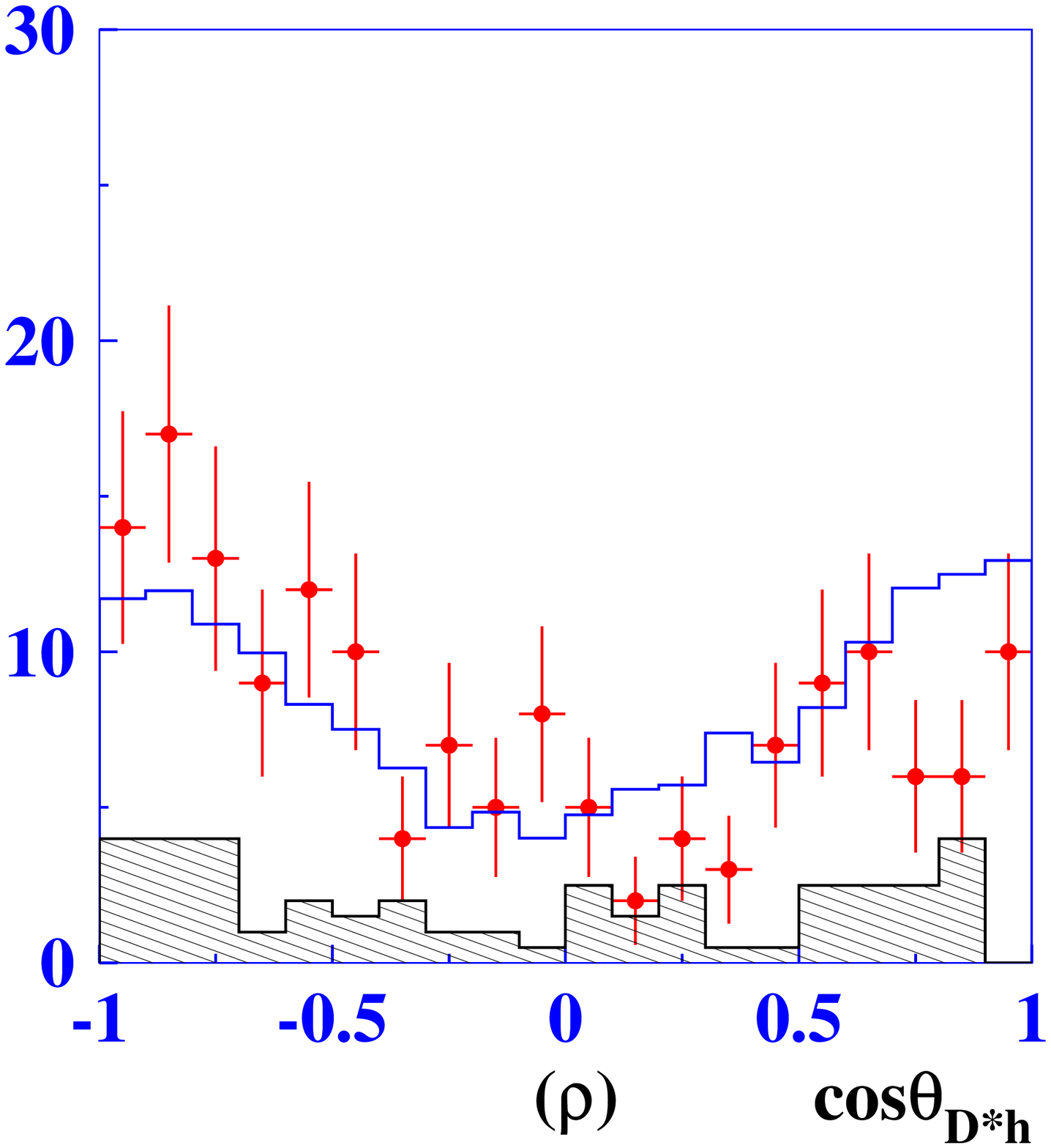}&
\includegraphics[height=5 cm]{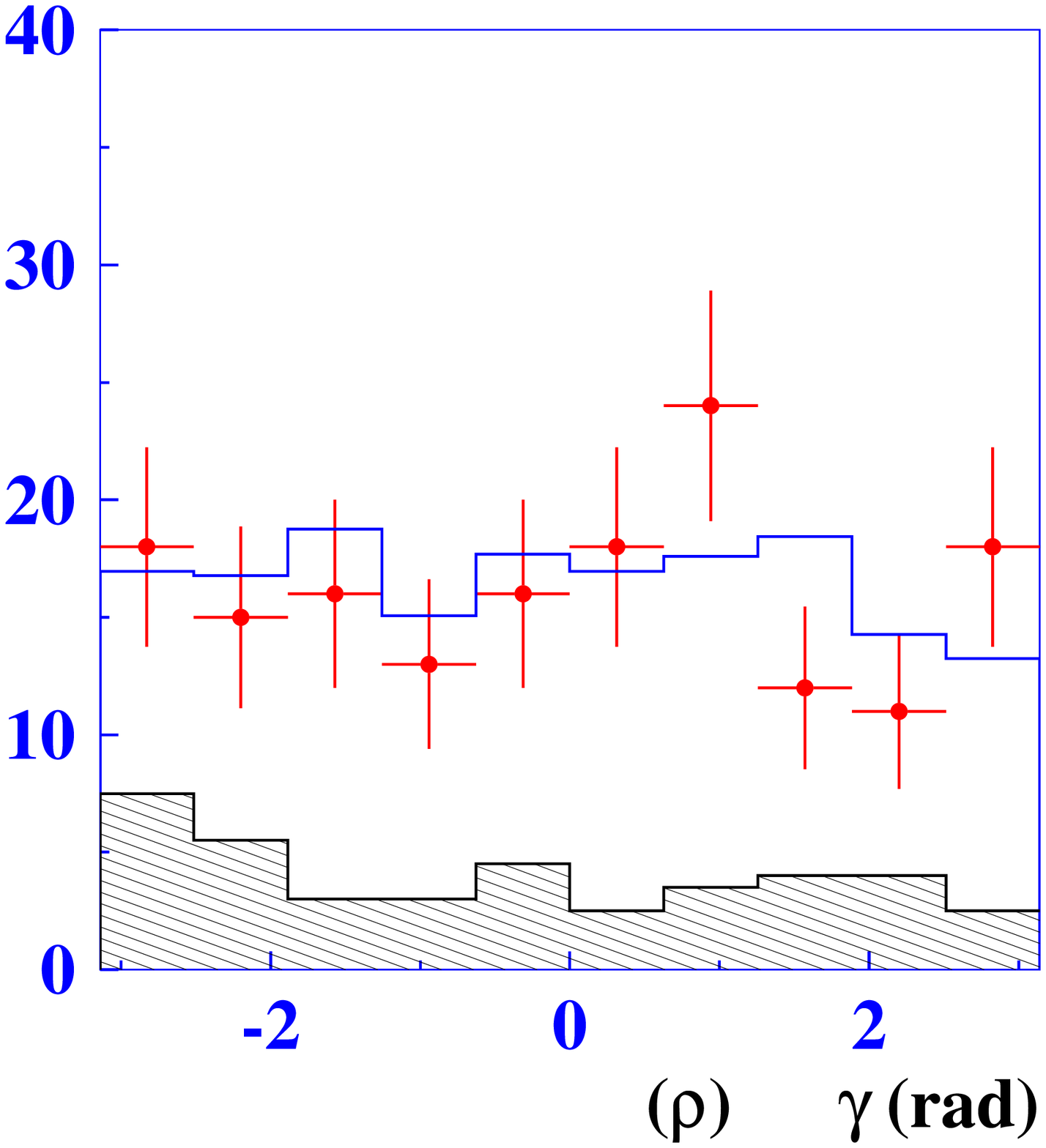}\\
\includegraphics[height=5 cm]{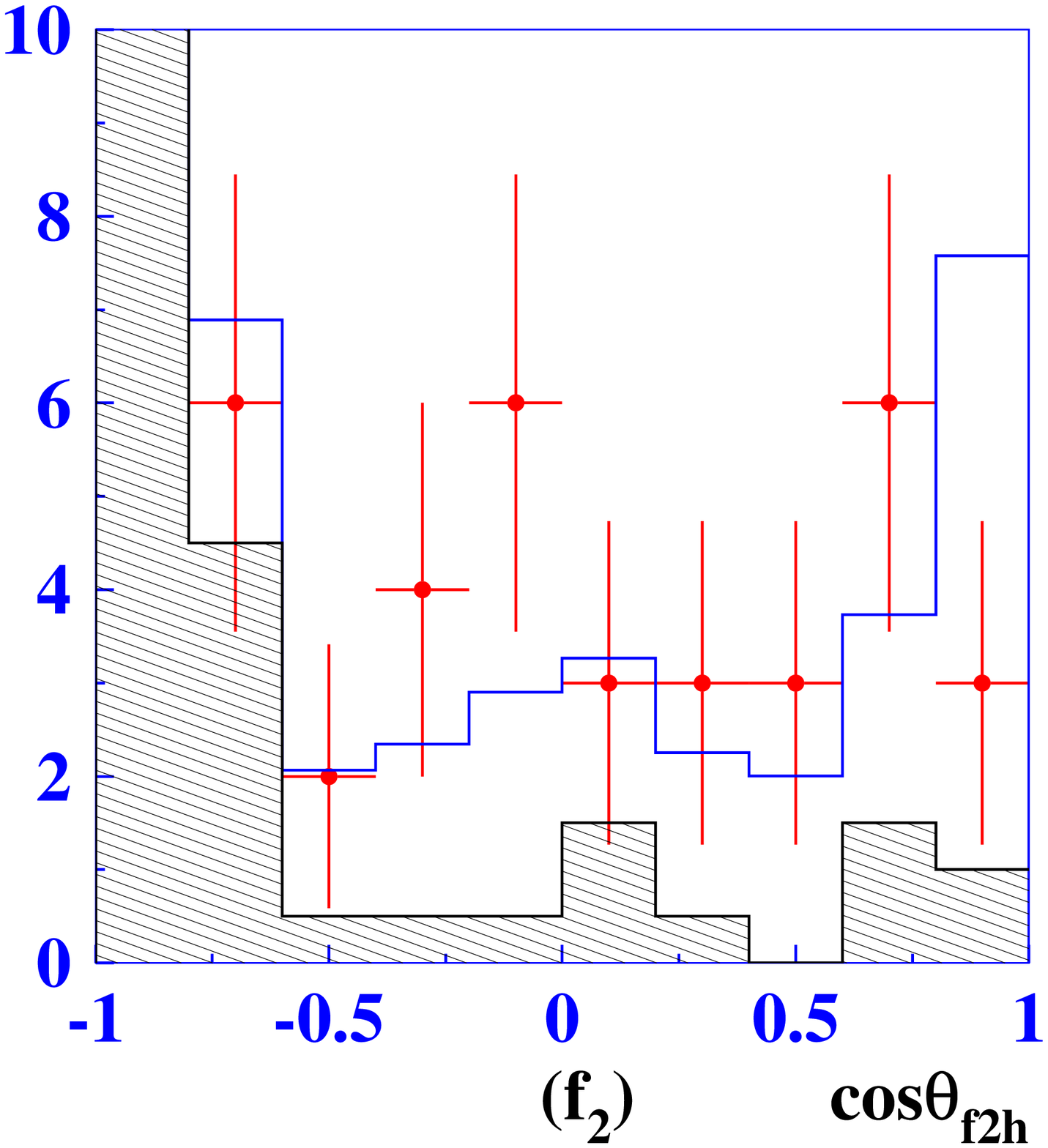}&
\includegraphics[height=5 cm]{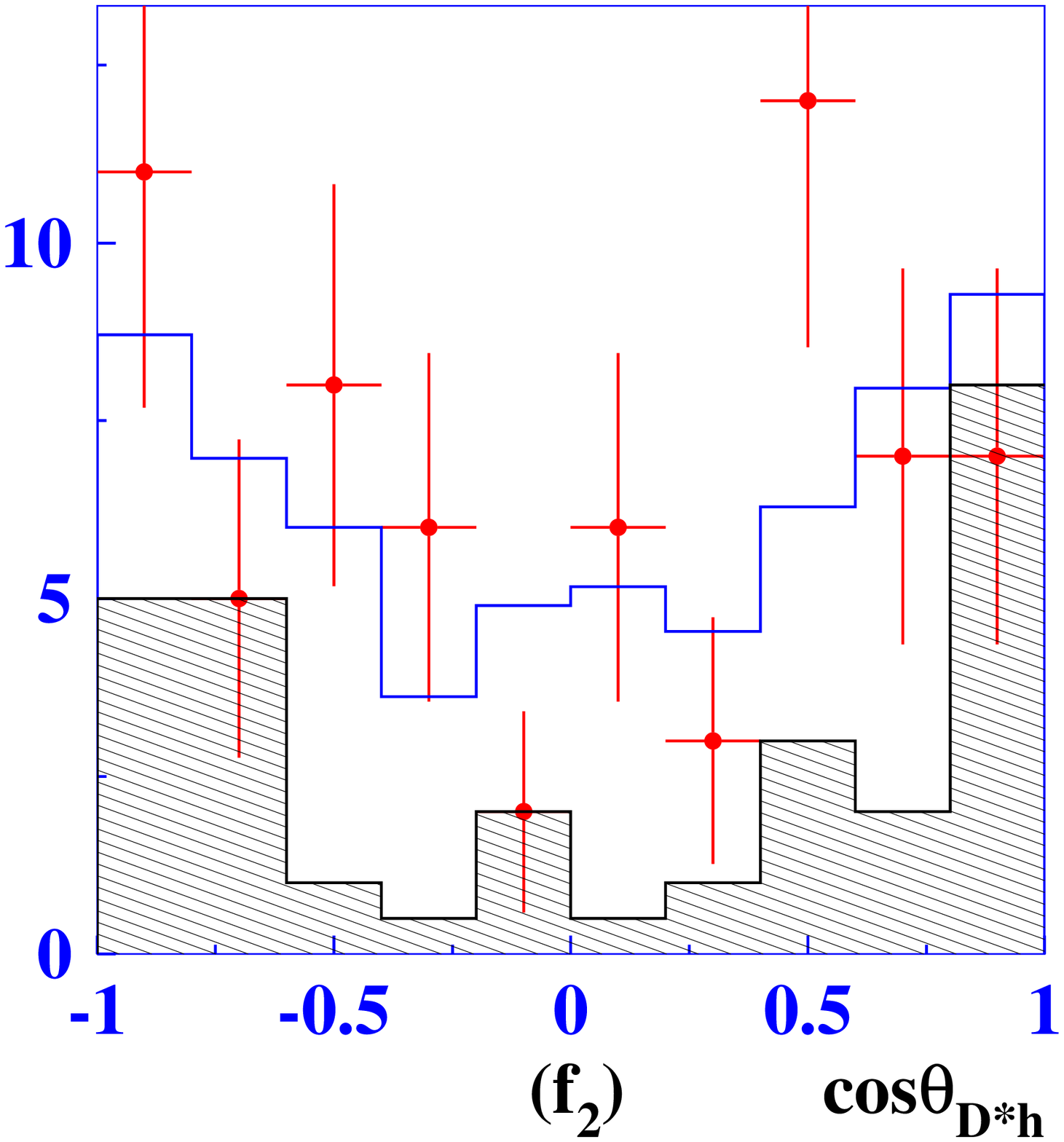}&
\includegraphics[height=5 cm]{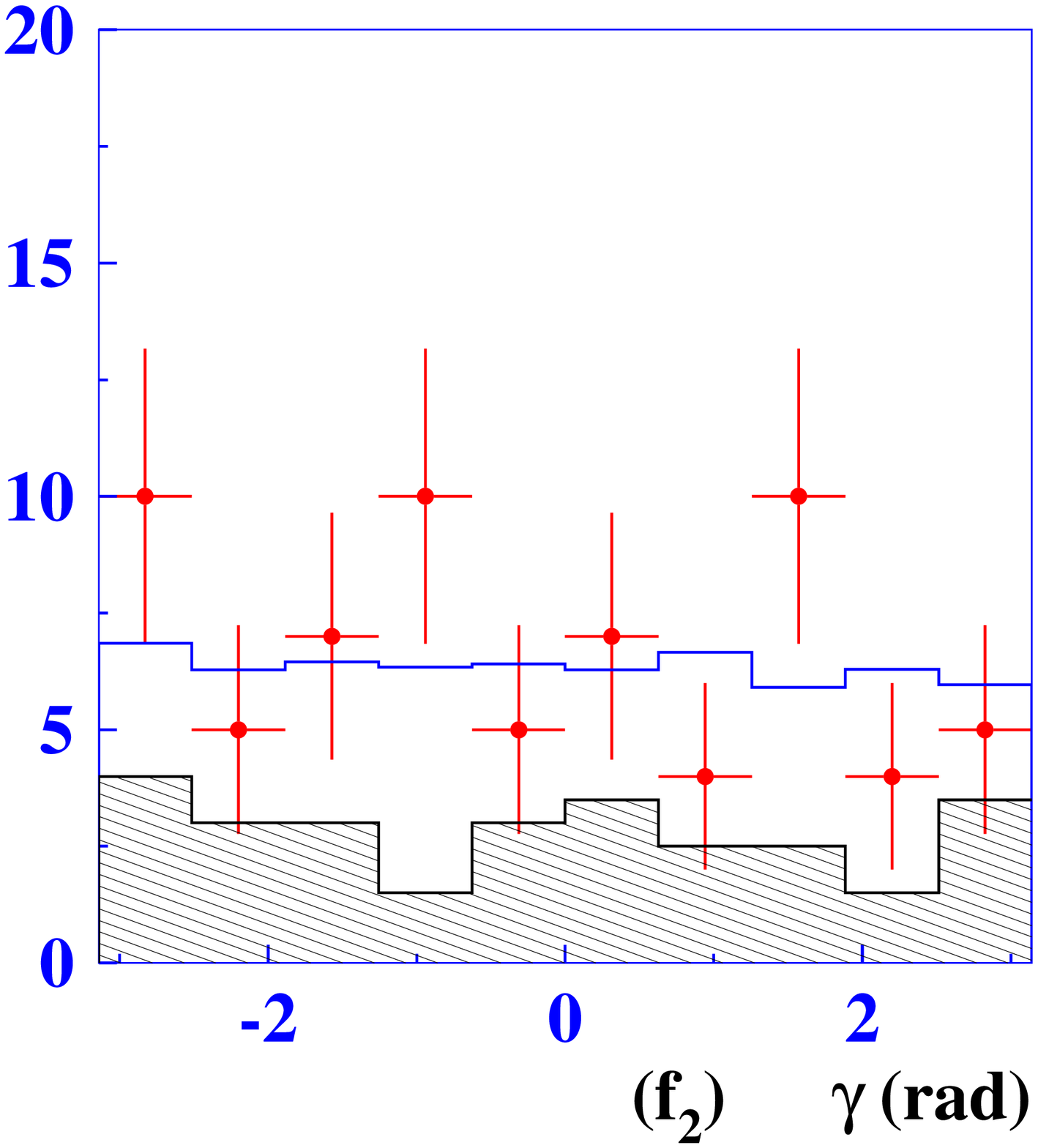}\\
\vspace*{-10 cm} & \\
{\bf\large \hspace*{-1cm} a)}&{\bf\large \hspace*{-1cm} b)}&{\bf\large \hspace*{-1cm} c)}\\
\vspace*{4 cm} & \\
{\bf\large \hspace*{-1cm} d)}&{\bf\large \hspace*{-1cm} e)}&{\bf\large \hspace*{-1cm} f)}\\
\vspace*{3 cm} & \\
\end{tabular}
\end{center}
\caption{Helicity distribution for  the $\rho$ region: (a)-(c)
  $|M_{\pi\pi}-.80|<0.20\,GeV/c^2$ and the $D_2^*$ region:
  $|M_{\pi\pi}-1.225|<0.125\,GeV/c^2$ (d)-(f).
(a), (d)  $\cos\theta$ -- helicity angle of $D^{**}$; 
(b), (e)  $\cos\alpha$ -- helicity angle of $D^{*}$; 
(c), (f)  azimuthal angle $\gamma$.experimental events (points) and for 
fast MC simulation (histogram). The hatched distribution shows 
the background distribution from the $\Delta E$ sideband region with 
a proper normalization. 
}
\label{f:dpp_hel_s}
\end{figure}

The helicity of the $\pi\pi$  system and the $D^*$ as well as the  azimuthal
angle $\gamma'$  are plotted in Fig.~\ref{f:dpp_hel_s} for  the $M_{\pi\pi}$ range
of the $\rho$ and the $f_2$.

The branching ratio for the $f_0(600)$ production is comparable with zero:
$
{\cal B}(\bar{B}^0\to f_0 D^0){\cal B}(f_0\to \pi^+\pi^-)=
(0.17\pm0.11\pm0.10^{+0.18}_{-0.05})\times10^{-4}.
$ This contribution can also  be regarded  as some nonresonance background. 

The branching ratios observed for the $\rho$ and the $f_2$ are as follows:
$$
{\cal B}(\bar{B}^0\to \rho^0 D^{*0})=
(3.73\pm0.87\pm{0.46}^{+0.18}_{-0.08})\times10^{-4},
$$
$$
{\cal B}(\bar{B}^0\to f_2 D^{*0}){\cal B}(f_2\to \pi^+\pi^-)=
(1.05\pm0.37\pm0.34^{+0.45}_{-0.33})\times10^{-4}.
$$
Taking into account the branching fraction of ${\cal B}(f_2\to
\pi^+\pi^-)$ we obtain:
$$
{\cal B}(\bar{B}^0\to f_2 D^{*0})=
(1.86\pm0.65\pm0.60^{+0.80}_{-0.52})\times10^{-4}.
$$

In spite of the limited statistics, we can determine the 
contribution of different polarization amplitudes for $\rho$ and
$f_2$:
$$
a^{\rho}_{0}=0.73\pm0.06\pm0.10\pm0.09
$$
$$
a^{\rho}_{||}=0.20\pm0.06\pm0.03\pm0.10
$$
$$
a^{\rho}_{\perp}=0.07\pm0.04\pm0.05_{-0.03}^{+0.19}
$$

$$
a^{f_2}_{0}=0.62\pm0.14\pm0.25\pm0.24
$$
$$
a^{f_2}_{||}=0.30\pm0.14^{+0.07+0.09}_{-0.27-0.27}
$$
$$
a^{f_2}_{\perp}=0.08\pm0.08^{+0.21+0.21}_{-0.03-0.02}
$$

\subsection{Results and discussion}

\begin{figure}[h]
\begin{center}
\begin{tabular}{ccc}
\includegraphics[width=5 cm]{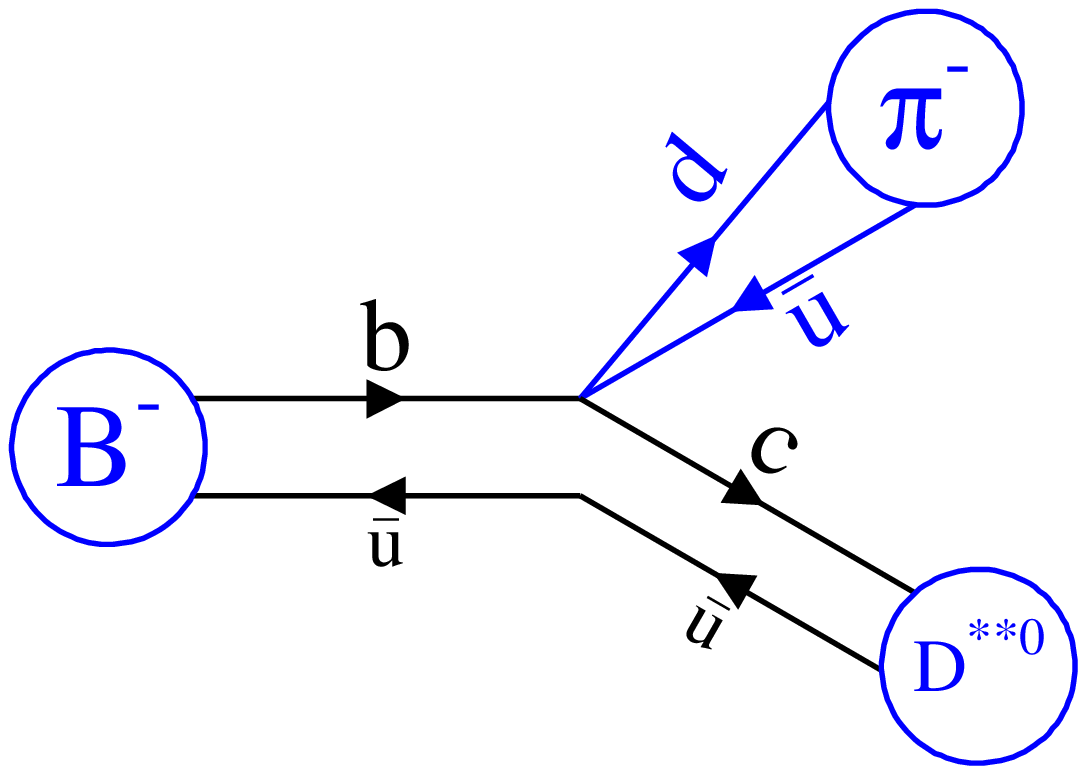}&
\includegraphics[width=5 cm]{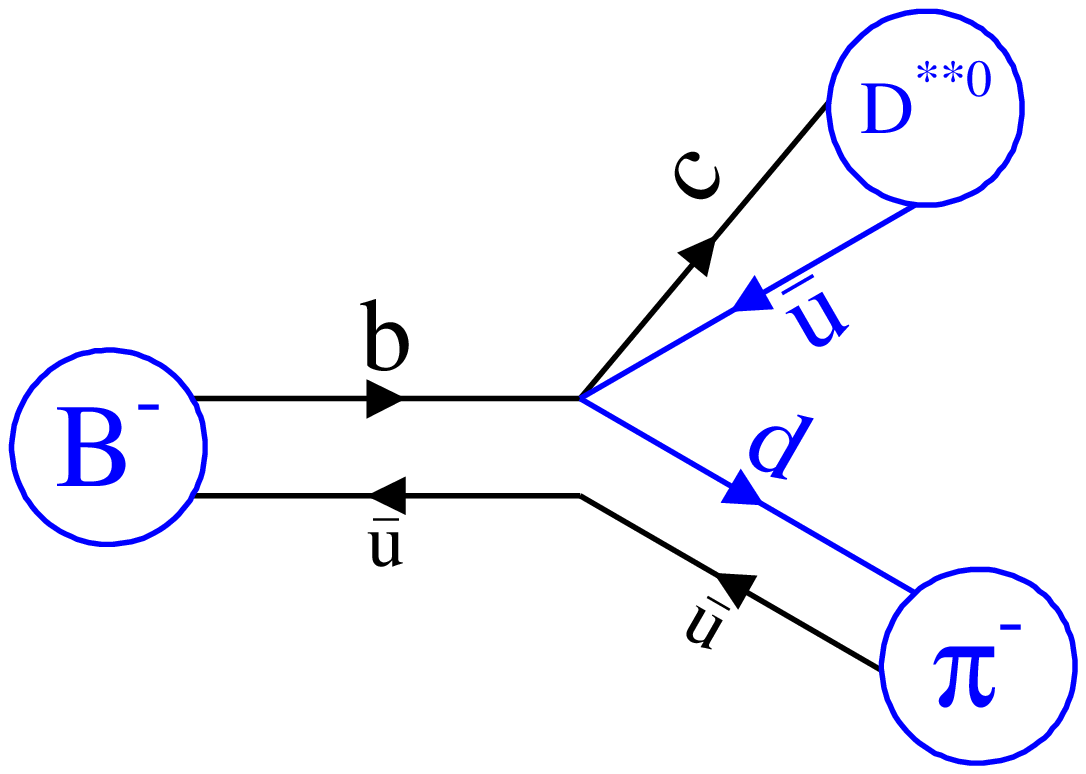}&
\includegraphics[width=5 cm]{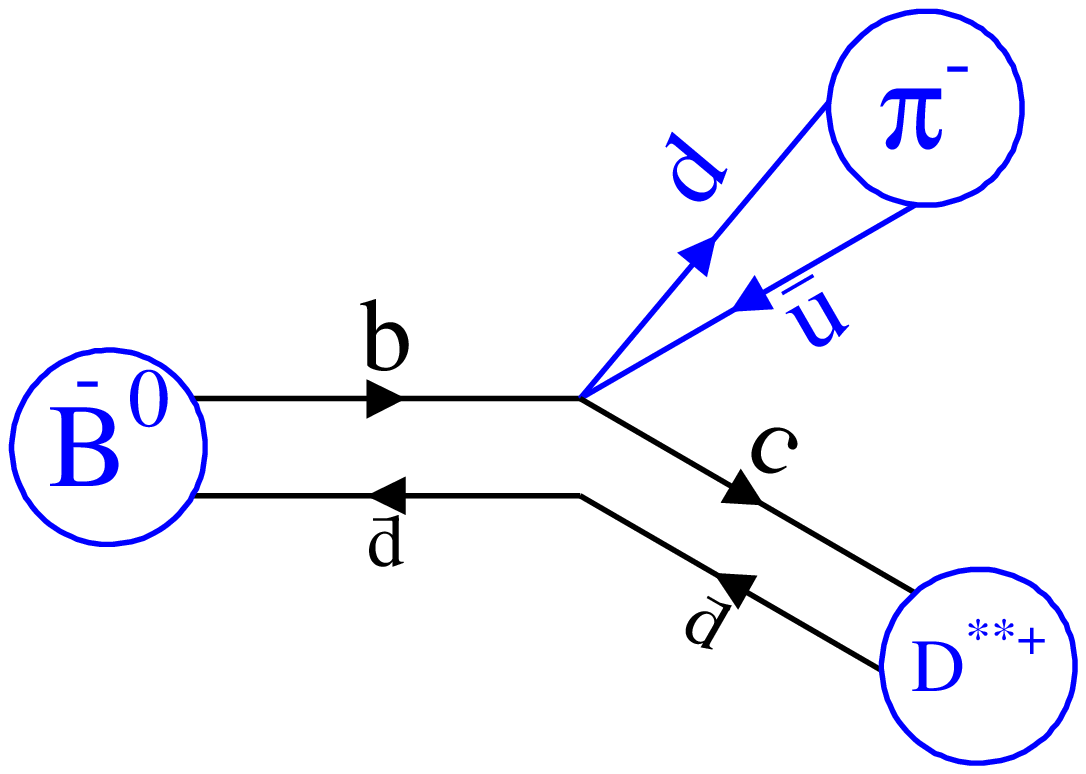}\\
\vspace*{-4 cm} & \\
{\bf\large \hspace*{-1cm} a)}&{\bf\large \hspace*{-1cm} b)}&{\bf\large
  \hspace*{-1cm} c)}\\ 
\vspace*{3 cm} & \\
\end{tabular}
\caption{Feynman diagrams for charged (a),(b) and neutral (c) $B$ decays.} 
\label{f:fd1}
\end{center}
\end{figure}

The branching-fraction products obtained for the narrow $(j=3/2)$
resonances are similar to
those obtained in the case of charged $B$
decays as shown in Table~\ref{zak}.
\begin{table}
\begin{tabular}{|c|c|c|}
\hline
& Neutral $B$ & Charged $B$~\cite{mybelle}\\
\hline
$
{\cal B}(\bar{B}\to D^{*}_2\pi^-){\cal B}(D_2^{*}\to
D^{*}\pi)
$
&
$
(2.45\pm0.42^{+0.35+0.39}_{-0.45-0.17})\times10^{-4}
$
&
$
(1.8\pm0.3\pm0.3\pm0.2)\times10^{-4}
$\\
$
{\cal B}(\bar{B}\to D_1\pi^-){\cal B}(D_1\to
D^{*}\pi)
$
&
$
(3.68\pm0.60^{+0.71+0.65}_{-0.40-0.30})\times10^{-4}
$
&
$
(6.8\pm0.7\pm1.3\pm0.3)\times10^{-4}
$\\
$
{\cal B}(\bar{B}\to D^{*}_2\pi^-){\cal B}(D_2^{*}\to
D\pi)
$
&
$
(3.08\pm0.33\pm0.09^{+0.15}_{-0.02})\times10^{-4}
$
&
$
(3.4\pm0.3\pm0.6\pm0.4)\times10^{-4}
$\\
\hline
$
{\cal B}(\bar{B}\to D'_1\pi^-){\cal B}(D'_1\to
D^{*}\pi)
$
&
$
<0.7\times10^{-4} \rm ~at~90\,\%~C.L.
$
&
$
(5.0\pm0.4\pm1.0\pm0.4)\times10^{-4}
$\\
$
{\cal B}(\bar{B}\to D^{*}_0\pi){\cal B}(D_0^{*}\to D\pi)
$
&
$
<1.2\times10^{-4} \rm~at~90\,\%~C.L.
$
&
$
(6.1\pm0.6\pm0.9\pm1.6)\times10^{-4}
$\\
\hline
\end{tabular}
\caption{Comparison of branching-fraction products for neutral
  and charged $B$ decays.}
\label{zak}
\end{table}
The measured values of the branching fractions of the broad resonance 
$D^{*+}_0$ and $D'^{+}_1$ production in neutral $B$ decay are, however, 
significantly lower than those for charged $B$ decays.
One possible explanation for this phenomenon is that 
for charged $B$ decay to $D^{**}\pi$, the amplitude
receives contributions from both tree and the color
suppressed diagrams as shown in Fig.~\ref{f:fd1}.
Production of $D^{**}$ via tree-diagrams is described
by the Isgur-Wise functions $\tau_{1/2}$ and  $\tau_{3/2}$.
According to the sum rule~\cite{QCDSR1,QCDSR}, $\tau_{1/2}\ll \tau_{3/2}$ and
one would expect the suppression of the broad state production.
For the color suppressed diagrams, however, $D^{**}$'s are produced by
another 
mechanism and the amplitudes are characterized by the constants $f_{D(3/2)}$ and 
$f_{D(1/2)}$ and  $f_{D(3/2)} \ll f_{D(1/2)}$.
The production of the broad resonances 
$D^{*0}_0$ and $D'^{0}_1$ in charged $B$ decay is probably 
amplified by
the color suppressed amplitude. 

\section{Conclusion}

A study of neutral $B$ to $D^0\pi^+\pi^-$ and $D^{*0}\pi^+\pi^-$ decays
has been presented.  
We have measured the total branching fractions of the three-body decays:
${\cal B}(\bar{B}^0\to D^0\pi^+\pi^-)
=(1.07\pm0.06\pm0.10)\times10^{-3}$ 
and 
${\cal B}(\bar{B}^0\to
D^{*0}\pi^+\pi^-)=(1.09\pm0.08\pm0.16)\times10^{-3}$
and the two-body decay:
${\cal B}(\bar{B}^0\to D^{*+}\pi^-)=(2.30\pm0.06\pm0.19)\times10^{-3}$.

The dynamics of these three-body decays has  been studied.
The $D^0\pi^+\pi^-$ final state is described by the  
production of $D^*_2\pi^-$  with
subsequent decays of $D^{*}_2\to D\pi$ and $D\rho,~Df_2$ and a 
broad scalar ($\pi\pi$) structure.
From a Dalitz distribution analysis we have obtained  the branching fraction product
for $D^{*+}_2$: 
$$
{\cal B}(\bar{B}^0\to D^{*+}_2\pi^-)\times B(D_2^{*+}\to D^{0}\pi^+)=(3.08\pm0.33\pm0.09^{+0.15}_{-0.02})\times10^{-4}.
$$
The values obtained for the mass and width of the tensor meson $D^{*+}_2$ are:
$$
M_{D^{*+}_2}=(2459.5\pm2.3\pm0.7^{+4.9}_{-0.5}) {\rm MeV}/c^2,~~\Gamma_{D^{*+}_2}=(48.9\pm5.4\pm4.2\pm1.9){\rm MeV}.
$$

The upper limit for the contribution of the  scalar $D^{*+}_0$ meson
assuming its mass and width of $D_0^0$ is:
$$
{\cal B}(\bar{B}^0\to D^{*+}_0\pi^-)\times B(D_0^{*+}\to
D^{0}\pi^+)<1.2\times10^{-4} \rm~at~90\,\%~C.L..
$$

The branching fractions for $D\rho$ and $Df_2$ productions have been measured:
$$
{\cal B}(\bar{B}^0\to \rho^0 D^0)=
(2.91\pm0.28\pm{0.33}^{+0.08}_{-0.54})\times10^{-4}
$$
$$
{\cal B}(\bar{B}^0\to f_2 D^0)=
(1.95\pm0.34\pm0.38^{+0.32}_{-0.02})\times10^{-4}.
$$
The last result represents a first observation.

The $D^*\pi\pi$ final state is described  by the 
production of $D^*_2\pi$ and $D_1\pi$ with
subsequent decays of $D^{**}\to D^*\pi$ and  $D^*\rho,~D^*f_2$.
From a phase space analysis, we obtain the branching fractions product
for $D^{**0}$: 
$$
{\cal B}(\bar{B}^0\to D_1^+\pi^-)\times B(D_1^+\to D^{*0}\pi^+)=(3.68\pm0.60^{+0.71+0.65}_{-0.40-0.30})\times10^{-4},
$$
$$
{\cal B}(\bar{B}^0\to D^{*+}_2\pi^-)\times B(D_2^{*+}\to D^{*0}\pi^+)=(2.45\pm0.42^{+0.35+0.39}_{-0.45-0.17})\times10^{-4}
$$
and set an upper limit on the production of the broad $D'_1$ resonance:
$$
{\cal B}(\bar{B}^0\to D'^{+}_1\pi^-)\times B(D'^{+}_1\to
D^{*0}\pi^+)<0.7\times10^{-4} \rm~at~90\,\%~C.L..
$$
For  the $D_1$ meson mass and width we obtain the following values:
$$
M_{D^{+}_1}=(2428.2\pm2.9\pm1.6\pm0.6)\, {\rm MeV}/c^2,~\Gamma_{D^{+}_1}=(34.9\pm6.6^{+4.1}_{-0.9}\pm4.1)\, \rm{MeV}.
$$

The branching fraction of $D^*\rho$ and $D^*f_2$ has been measured:
$$
{\cal B}(\bar{B}^0\to \rho^0 D^{*0})=
(3.73\pm0.87\pm{0.46}^{+0.18}_{-0.08})\times10^{-4}
$$
$$
{\cal B}(\bar{B}^0\to f_2 D^{*0})=
(1.86\pm0.65\pm0.60^{+0.80}_{-0.52})\times10^{-4},
$$
These are the first measurements of these processes.
We also observe dominance of the longitudinal polarization amplitude
for $B\to D^*\rho$ and $B\to D^*f_2$.

All results are preliminary.

\section*{Acknowledgments}
We thank the KEKB group for the excellent operation of the
accelerator, the KEK Cryogenics group for the efficient
operation of the solenoid, and the KEK computer group and
the National Institute of Informatics for valuable computing
and Super-SINET network support. We acknowledge support from
the Ministry of Education, Culture, Sports, Science, and
Technology of Japan and the Japan Society for the Promotion
of Science; the Australian Research Council and the
Australian Department of Education, Science and Training;
the National Science Foundation of China under contract
No.~10175071; the Department of Science and Technology of
India; the BK21 program of the Ministry of Education of
Korea and the CHEP SRC program of the Korea Science and
Engineering Foundation; the Polish State Committee for
Scientific Research under contract No.~2P03B 01324; the
Ministry of Science and Technology of the Russian
Federation; the Ministry of Education, Science and Sport of
the Republic of Slovenia; the National Science Council and
the Ministry of Education of Taiwan; and the U.S.\
Department of Energy.


\begin{thebibliography}{120}
\bibitem{foot1}{The $z$ coordinate of the track is
defined as the $z$ coordinate of the track point closest to the beam in
the $r-\phi$ 
plane. The $z$  axis is opposite the positron beam direction.}
\end{thebibliography}
\end{document}